# Analysis of surface tension in terms of force gradient per unit area - Part I : A thought experiment using the principle of equivalence in fluid mechanics

*André Schiltz*
*Grenoble, France*

## Abstract

*Conventionally, surface tension is expressed as a force per unit length or as an energy per unit area.*
*In this paper, we propose a thought experiment that consists of replacing the surface tension with an equivalent force per unit area according to the principles of fluid mechanics.*
*Such a system of equivalent forces makes it possible to analyze the surface tension phenomenon in terms of surface stress gradient or in terms of energy per unit volume and allows us to rewrite the known equations by calculating the equilibrium of forces in the stationary state. These new equations will be applied to known phenomena such as meniscus, capillary tube, Wilhelmy blade and equilibrium of drops and semi-drops.*

## I Introduction

Surface tension is a physicochemical phenomenon related to molecular interactions at the interface between a liquid and air, a solid or another liquid. This phenomenon is responsible for the effects of capillarity, the formation of a meniscus at the edge of a glass, the formation of drops of water, the fact that insects seem to walk on the water, etc.
Classically, the surface tension ‘γ’ is defined thermodynamically either as a force per unit length in [N/m], or as an energy per unit area in [J/m$^2$].
In this paper, we propose a thought experiment by mechanically defining equivalent stress gradients that allow us to interpret surface tension effects in terms of force gradient per unit area in [N/m$^2$] or in terms of energy gradient per unit volume in [J/m$^3$].

### 1.1 Law of Laplace and equation of Young-Laplace (1804)

In 1804, Pierre-Simon Laplace[1] took up the observations of Thomas Young and published his ‘*Theory of capillary action*’ where he presented a mathematical analysis of the average curvature of a surface, which makes it possible to calculate the pressure difference in a drop of water or a bubble with respect to the atmosphere. The pressure difference is greater as the drop or bubble is smaller and it is proportional to the surface tension ‘γ’ according to the so-called equation of Young-Laplace :

$$\Delta P = \frac{2\ \gamma}{R} \tag{1}$$

Where :
ΔP : is the pressure difference at the interface [N/m$^2$]
R : is the radius of the spherical drop [m]
γ : is the surface tension [N/m] or [J/m$^2$]

The Young-Laplace’s equation is obtained by dividing the projection of the surface tension ‘γ ’ along the perimeter ‘2ΠR’ by the surface of the median plane 'ΠR$^2$’, that is : $\Delta P = (2\ \gamma /R)$.

The main limitation of the Young-Laplace's equation comes from the fact that the pressure theoretically tends towards infinity when the radius is zero. To this day, this equation is still widely used to calculate the effects of capillarity, as for example in the case of rising liquid in a capillary tube, according to the so-called Jurin's law[2].

### 1.2 The Young-Dupré equation (1805)

Regarding the equilibrium of meniscus and that of a drop of water on a plane, we generally refer to the Young-Dupré law[3]. As described in **Figure 1**, according to this law, the three surface tensions associated with the three interfaces must be in equilibrium for the triple line to stop.
The vector sum of the three tensions projected onto the surface is in principle zero as in the Young-Dupré equation :

$$\gamma_{LV} \cos(\theta) = \gamma_{SV} - \gamma_{SL} \qquad (2)$$

Where :
- $\theta$ : is the contact angle of the drop or of the meniscus
- $\gamma_{LV}$ : is the liquid-vapor surface tension
- $\gamma_{SL}$ : is the solid-liquid surface tension
- $\gamma_{SV}$ : is the solid-vapor surface tension

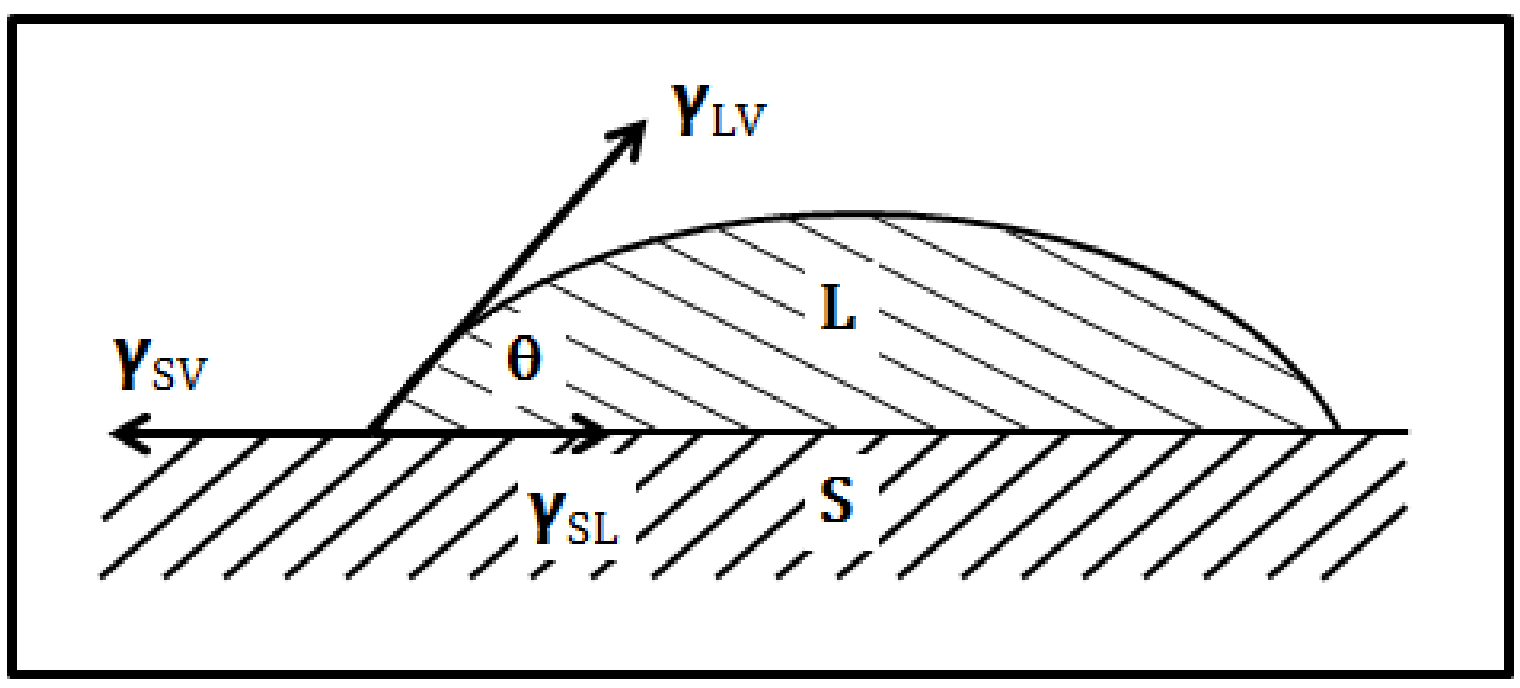

**Figure 1 – The Young-Dupré Equation. The case of a drop of water**

To solve this equilibrium equation with three unknowns, experimenters usually measure the contact angle '$\theta$' at the solid-liquid-vapor triple point, they use the value of the liquid-vapor surface tension '$\gamma_{LV}$' measured using a tensiometer and they calculate the solid-liquid surface tension by extrapolation using the Zisman and Fox method[4].
In the case of the meniscus, the diagram in **Figure 2** describes the balance between the three surface tensions '$\gamma_{LV}$', '$\gamma_{SL}$' and '$\gamma_{SV}$' following the Young-Dupré equation. Note that the solid-vapor surface tensor is upwards, while the liquid-vapor and solid-liquid surface tensors are downwards[5].
As we will see in the examples of the capillary tube and the Wilhelmy plate, most experimenters consider only the surface tension '$\gamma_{LV}$' and the '$\theta$' angle, this angle being linked to the two other tensors by the Young-Dupré equation.

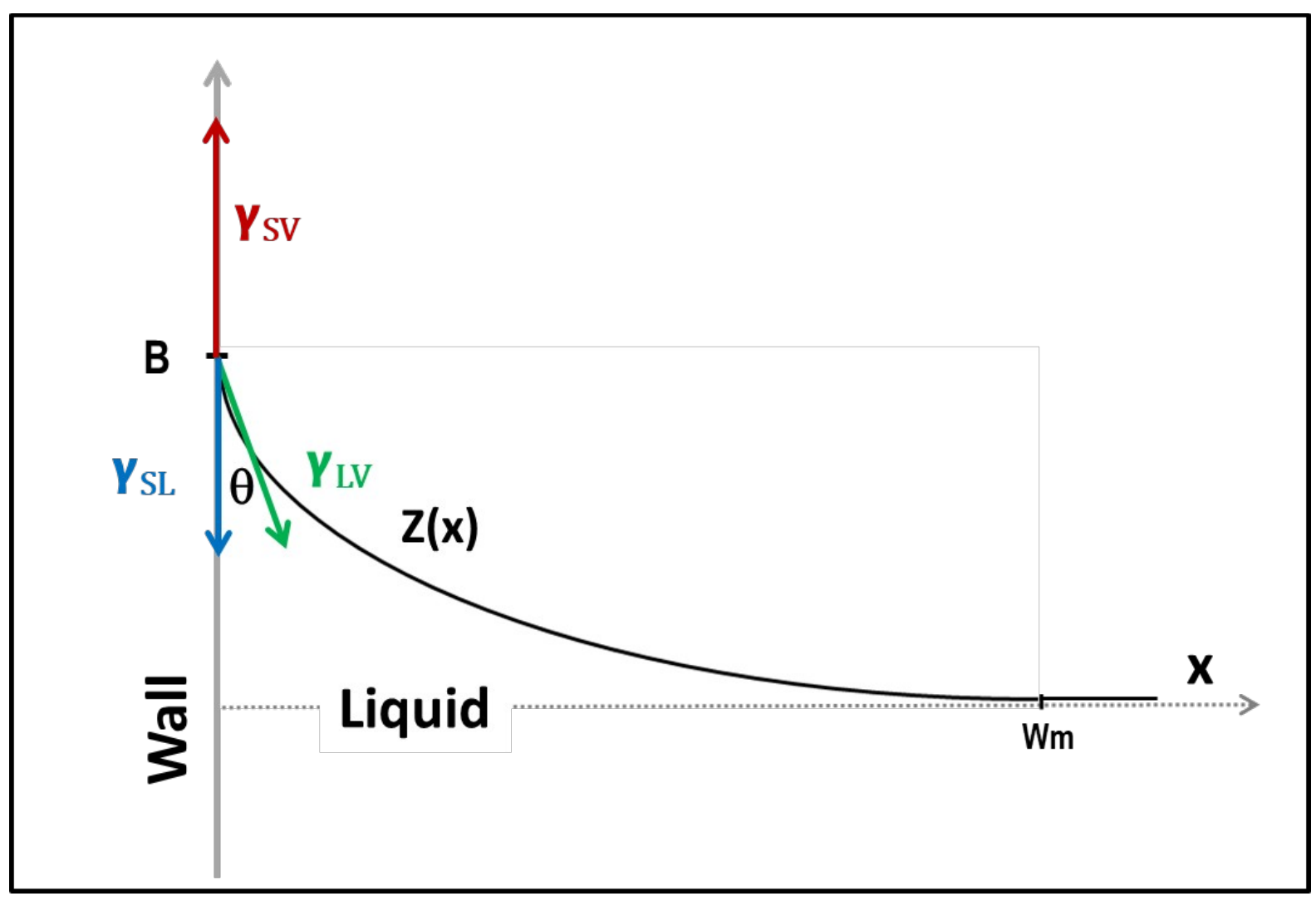

**Figure 2 – Schematic of a meniscus with surface tensions and contact angle following Young-Dupré.**

*Meniscus equation using capillary length*

Most authors consider that the profile of the meniscus has very approximately the shape of a circular arc, but some authors are more realistic and use an exponential, such as D. Vella et al.[6] and F. Elie[7].
Interestingly, F. Elie balances the hydrostatic pressure variation with the Laplace pressure by using the general equation for the radius of curvature of the meniscus surface, i.e., '$R = - \partial^2 z/\partial x^2$' and writes the following differential equation : $\partial^2 z/\partial x^2 - \rho g z/\gamma = 0$.
Using the definition of the capillary length, $L_c = \sqrt{\frac{\gamma}{\rho g}}$ , he rewrites it as : $\partial^2 z/\partial x^2 - z/L_c^2 = 0$.
This differential equation has an exponential solution such as :

$$Z(x) = Z_0\, e^{-\, x/Lc} \tag{3}$$

According to this equation, the profile of the meniscus thus defined would depend only on the two parameters '$Z_0$' and '$L_c$'. The authors point out that this is an approximation.
Note that this equation is exponential, like the ones we derive in this article.

### 1.3 Energy aspect

The surface tension 'γ' can be considered either as a force per unit length in [N/m] or as an energy per unit area in [J/m$^2$] :

a) As a force per unit length, '**γ**' is the force in Newtons needed to stretch the interface surface by one meter.
b) As an energy per unit area, '**γ**' is the energy in Joules needed to increase the interface area by one square meter.

P-G. de Genes et al.[8] consider that surface tension is a physico-chemical phenomenon related to the increase in energy at the interface between two fluids or at the fluid/air interface.
In a liquid, the interaction forces between molecules are in equilibrium and their resultant is zero. At the liquid-air interface, equilibrium cannot be maintained, and surface molecules are attracted inwards, creating a force known as « surface tension ».
Considering that molecules at the surface have fewer interactions with their neighbors than in the volume, and taking into account the Brownian effect of molecular motion[6], they estimate that if 'U' is the cohesion energy per molecule in volume, the energy at the surface should be in the order of '*U/2*'.

If '$a$' is the size of a molecule, '$a^2$' is the area exposed to the surface, and it can therefore be assumed that the surface tension 'γ', which measures the loss of energy at the surface, is worth '$\gamma \approx U/2a^2$'.
For common liquids, Van der Waals interactions predominate, and the thermal energy is calculated as ' $U \cong k_B.T$ ', where '$k_B$' is the Boltzmann constant, which gives a value of 'γ' close to $20\ mJ/m^2$ at 25°C. In the case of water, where : γ ≈ 0.072 N/m, it is explained that the surface tension measured is greater because hydrogen bonds are predominant.

### 1.4 Comments on the thermodynamic vision of interface

Surface tension is thermodynamically defined as an amount of energy per unit area. This means that all the energy is supposed to be concentrated at the interface on a mathematical surface with no thickness.
Regarding the question of the thickness of the molecular layer on the surface of the fluid, there are several schools of thought : (i) J. W. Gibbs considers the surface as a mathematical surface without thickness whereas (ii) J. D. Van der Waals and H. Bakker assigned a thickness of the size of the Van Der Waals interactions[9].
In their 2002 bibliographic review, L. J. Michot et al.[10] subscribe to this molecular vision, pointing out that « Whatever the support, it is now clearly evidenced that structural perturbations are limited to distances lower than 10–15 Å from the interface.» , which is 3 to 5 times the size of a water molecule.
Thus, the question of energy distribution at the interface seems to be resolved, as no structural perturbations beyond a few molecular diameters are observed.
However, if we have a look at the meniscus in a glass of water, we can see that it is not nanometric but millimetric in size. The classical explanation is that surface forces act tangentially on the liquid surface, bending it over several millimeters and causing the liquid to move below the meniscus line.

## II Mechanical vision and definition of surface stress gradients

### 2.1 Mechanical concept of equivalent forces and thought experience

In fluid mechanics, to account for deformations induced in a volume by a normal or a tangential force, this force can be mathematically replaced by a system of equivalent (tangential, normal and/or shear) forces, respecting the laws of conservation of mass and momentum, provided of course that these forces have the same effect on the volume.
This principle is illustrated in **Figure 3.a**, where we can see that the deformation of volume 'A' into volume 'B' can theoretically come from either longitudinal elongation or transverse compression.

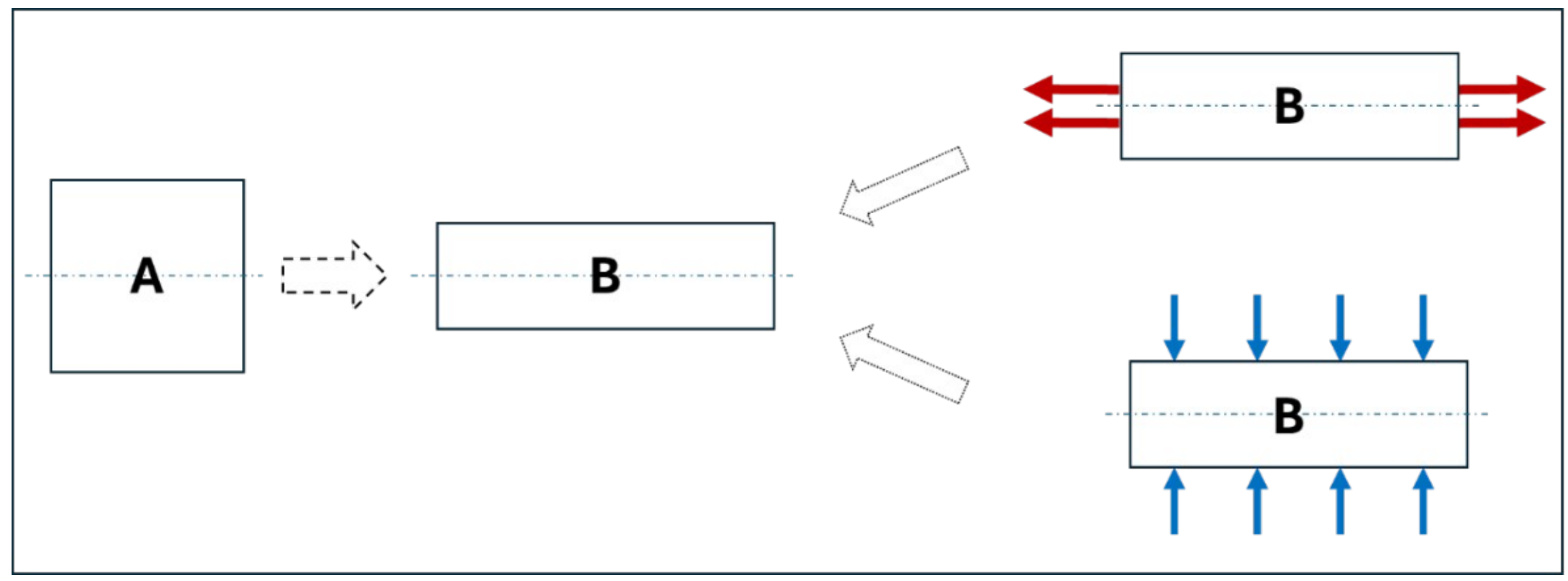

**Figure 3a – Principle of mechanical equivalence**

**Thought experiment : mechanical equivalence model**
Similarly, the following experience of thought can be suggested :
Instead of assuming that the meniscus in **Figure 2** is derived from the Young-Dupré analysis, i.e. the effect of tangential surface tension forces, we can theoretically assume that the same result could be achieved with equivalent forces as shown in **Figure 3.b**.

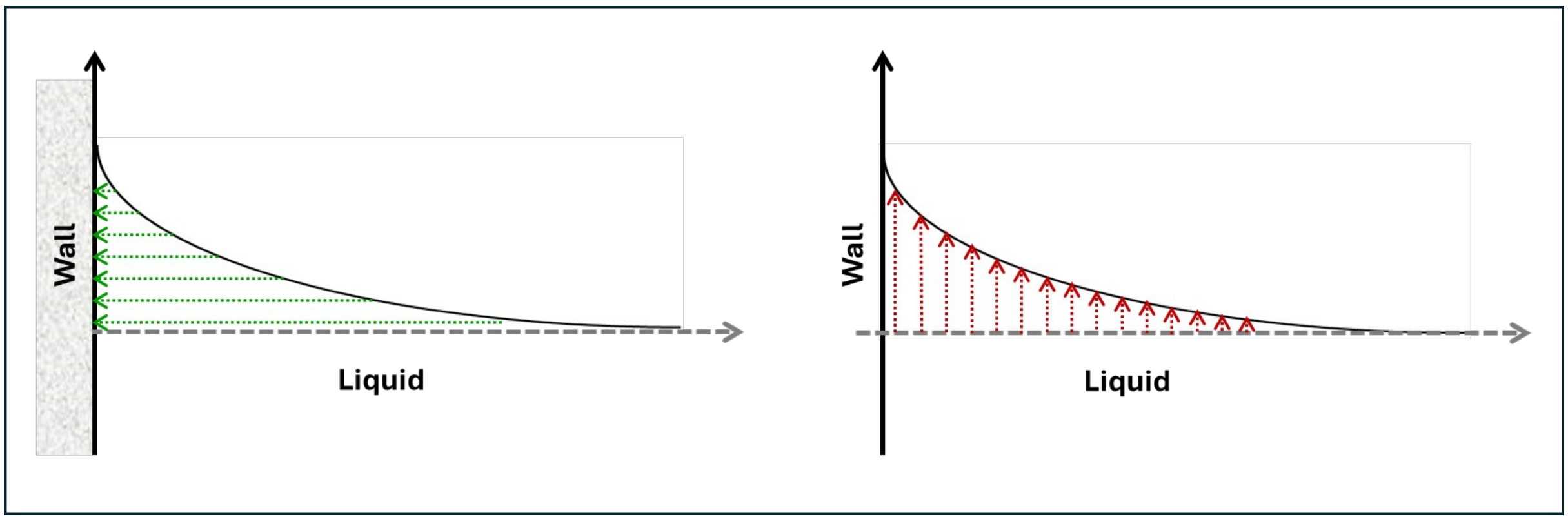

**Figure 3b – Theoretical scheme of equivalent forces that can lead to meniscus formation**

In this article, we will therefore mathematically replace conventional surface tensions with an equivalent stress gradient and first apply this concept to the case of the meniscus.

### 2.2 Mechanical analysis of forces in meniscus

The meniscus case is ideal for several reasons : The surfaces of the liquid and the solid are large compared to the dimensions of the meniscus and can be considered as "almost infinite". There are no horizontal or vertical limits imposed by the volume of the liquid or the dimensions of the solid. The meniscus profile can extend horizontally until it is damped at the surface of the liquid, and it can climb vertically to its maximum along the solid vertical wall, as shown schematically in **Figure 2**. We can therefore assume, as in the Young-Dupré equation, that the different forces are in equilibrium.
At the liquid-air interface, it is generally assumed that « the molecules of the surface are attracted towards the interior creating a force called surface tension »[5], and the surface tension (in [$Nm^{-1}$]) is defined tangential to the surface.
In this paper, we will consider that this 'attraction of molecules from the surface to the inside' or 'repulsion of molecules from the surface to the outside' creates equivalently a stress gradient of compression (in [$Nm^{-2}$]) perpendicular to the surface.
In the same way, we can consider that at the solid-liquid interface, the attraction of liquid molecules towards the solid wall creates equivalently a stress gradient parallel to the wall (in [$Nm^{-2}$]).
Thus, instead of the classical surface tensors, we will define stress gradients or force gradients per unit area. Such a mathematical construction is not insignificant because surface tension vectors are, by definition, tangential vectors to the surface, whereas the gradients defined here can be either perpendicular or parallel to the surface depending on the nature of the forces at the interface.

According to our hypothesis, in the meniscus case, **Figure 4a** shows repulsion forces at the liquid-vapor interface, which are perpendicular to the surface, and attraction forces at the solid-liquid and solid-vapor interface, which attract molecules to the solid wall.
The resulting force pushes the liquid molecules towards the wall and upwards to form the observable meniscus profile.
Equivalently, a surface stress gradient can be defined as in **Figure 4b**, which would theoretically be the gradient responsible for meniscus formation.
By definition, this equivalent stress gradient, called here the 'surface stress gradient', is in equilibrium with the pressure gradient, which represents a 'potential gravity energy per unit volume'.

In the specific case of the meniscus, the final deformation profile ‘Z(x)’ in **Figures 2** and **4** has the same shape as that of the final stress gradient. We will see further that this is no longer true for drops.

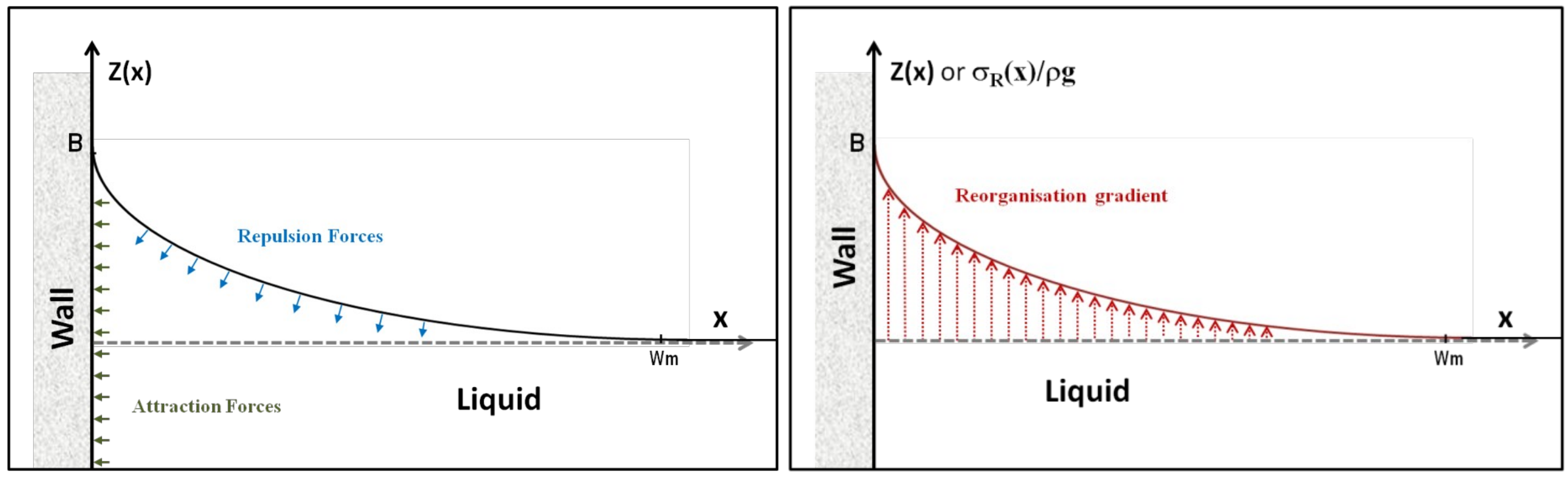

**Figure 4a – Scheme of theoretical attraction and repulsion forces at the interfaces of meniscus**
**Figure 4b – Scheme of equivalent stress gradient in meniscus**

Remember that in the case of water in a glass, the meniscus dimensions are of the order of 2 mm in height and more than 5 mm in attenuation length, according to the authors[11,12].

### 2.3 Surface energy and gravity potential energy

When the meniscus is in equilibrium, the only forces present are surface tension forces and gravity forces. We can therefore balance the resulting stress gradient or energy gradient with the pressure gradient or ‘gravity potential energy density per unit volume’.
Given the exponential shape of the meniscus, its profile ‘Z(x)’ can simply be written as :

$$Z(x) = B\, e^{-A x} \quad (4)$$

where :

B : is the meniscus height when ‘x=0’ [m]
A : is an attenuation constant [$m^{-1}$]

In the gravitational field and in stationary state, we can define a theoretical surface energy gradient ‘U(x)’ which equilibrates with potential gravity energy ‘$E_p(x)$’ as :

$$U(x) = E_p(x) = m\, g\, Z(x) \quad (5)$$

Representing these energies per unit volume, we can define a gravitational potential energy density per unit volume ‘$E_{pp}$’ and a surface energy per unit volume ‘u(x)’ such as :

$$u(x) = E_{pp}(x) = \rho\, g\, Z(x) \quad (6)$$

where :

$E_{pp}$ : is the potential gravity energy per unit volume or more simply the pressure gradient in [$kg\, m^{-1}\, s^{-2}$], [$Jm^{-3}$] or [$Nm^{-2}$]
m : is the mass in [kg]
ρ : is the density in [$kg\, m^{-3}$]
g : is the intensity of gravity in [$m\, s^{-2}$] or [N/kg]
Z(x) : is the meniscus observed profile in [m]

NB : Whereas hydrostatic pressure is defined as ‘p = ρ g h’, what we call here potential gravity energy per unit volume ‘$E_{pp}$ (x)’ is in fact a pressure gradient. Nevertheless, we will keep the ‘$E_{pp}$ (x)’ notation here to avoid any confusion with the global pressure.

Using equation (4), we obtain the expression of the surface energy gradient per unit volume :

$$u(x) = \rho\ g\ Z(x) = \rho\ g\ B\ e^{-A\ x} \qquad (7)$$

Note the final deformation profile ‘Z(x)’ has the same shape as the final stress gradient.
While these energies per unit volume have the dimension of a pressure ([$Nm^{-2}$]) or a stress, that is, of a force per unit area, equation (7) can be rewritten as :

$$\sigma_R(x) = \rho\ g\ B\ e^{-A\ x} \qquad (8)$$

where :

$\sigma_R(x)$ : is the resulting stress gradient [$Nm^{-2}$] or energy gradient per unit volume [$Jm^{-3}$].
B : is the meniscus height [m]
A : is an attenuation constant [$m^{-1}$]
$W_m$ : is the meniscus attenuation length [m]

In the following, we will preferably use the term 'stress gradient', since it can be considered as a force gradient per unit area ‘$\sigma_R$’ (in [$Nm^{-2}$]) knowing that it can also be considered as an energy gradient per unit volume (in [$Jm^{-3}$]).

**2.4 Definition of the surface stress gradients in the meniscus**

In this paper, we will assume that the global stress gradient ‘$\sigma_R$ (x)’ is the resultant of only two gradients : (1) a surface stress gradient at the liquid-vapor interface, which can be written as ‘$\tau_{LV}(x)$’ and (2) a resulting stress gradient at solid interfaces (solid-liquid and solid-vapor), which can be written as ‘$\sigma_S(x)$’.

(1) The gradient ‘$\tau_{LV}(x)$’ is related to the surface tension forces at the liquid-vapor interface and it can be matched with the classic liquid-vapor surface tension vector ‘$\gamma_{LV}$’. The definition of the gradient ‘$\gamma_{LV}$’ is not problematic because there is only one force gradient that is perpendicular to the surface.
(2) Logically, the gradient ‘$\sigma_S(x)$’ should be breakable down into two components : a solid-liquid gradient ‘$\sigma_{SL}(x)$’ and a solid-vapor gradient ‘$\sigma_{SV}(x)$’. These gradients should, of course, correspond to the classic surface tension vectors ‘$\gamma_{SL}$’ and ‘$\gamma_{SV}$’. However, for reasons inherent in the mechanical equivalence model, they cannot be discriminated against in the meniscus case, because the solid-liquid and solid-vapor forces are both perpendicular to the surface and the gradients are overlapping. We will return to this point in a later article.

Therefore, in this paper, we will maintain the concept of a resulting stress gradient ‘$\sigma_S(x)$’ representing the sum or difference of the solid-liquid and solid-vapor force gradients.
This simplification will not prevent us from rewriting the equations for common phenomena such as the meniscus, the capillary tube, the Wilhelmy plate, and the drops.
We will replace this simplified mechanical model with a phenomenological model in a later article.

**2.4.1 Definition of the solid-liquid interface stress gradient**

As described by many authors, the stress gradient at the liquid-vapor interface could result from the repulsion of molecules from the surface to the bulk.

The repulsive stress gradient at the liquid-vapor interface can be written as :

$$\tau_{LV}(x) = - \rho\, g\, \lambda\, e^{-\varepsilon x} \quad (9)$$

where :

$\tau_{LV}(x)$ : is the liquid-vapor interface stress gradient in $[Nm^{-2}]$ or $[Jm^{-3}]$
$\lambda$ : is a characteristic length of curvature [m]
$\varepsilon$ : is an attenuation constant $[m^{-1}]$

Thus, the repulsive stress gradient at the liquid-vapor interface '$\tau_{LV}(x)$' acts perpendicular to the surface of the liquid and decreases as one moves away from the surface.

### 2.4.2 Definition of the stress gradient at solid interfaces

As discussed above, we consider that the gradient related to surface tension forces at solid interfaces '$\sigma_S(x)$' results from the (solid-liquid) attractive forces of liquid molecules toward the wall as well as the (solid-vapor) attraction/condensation forces of vapor molecules above the meniscus.
It can be written as :

$$\sigma_S(x) = \rho\, g\, \delta\, e^{-\alpha x} \quad (10)$$

where :

$\sigma_S(x)$ : is the solid interface stress gradient in $[Nm^{-2}]$ or $[Jm^{-3}]$
$\delta$ : is a length of action parallel to the wall [m]
$\alpha$ : is an attenuation constant $[m^{-1}]$
$W_m$ : is the meniscus attenuation length [m]

At this stage, we have the following gradients :

(i) a resulting solid interface stress/deformation gradient '$\sigma_S(x)$', parallel to the wall and decreasing from the wall to the bulk
(ii) a liquid-vapor interface stress gradient due to repulsion '$\tau_{LV}(x)$', perpendicular to the liquid surface and decreasing from the surface to the bulk

We can now write the equation for the meniscus.

## 2.5 The meniscus equation

According to our hypotheses, the observable shape of the meniscus results from the interaction of the surface tension forces at the solid interfaces (related to gradient '$\sigma_S(x)$') and at the liquid-vapor interface (related to gradient '$\tau_{LV}(x)$'), which are in equilibrium with the gravity forces.
In this model, we assume that the initial meniscus profile '$Z_S(x)$' is first created by solid surface tension forces and that this profile is then distorted by liquid-vapor surface tension forces.
The final resulting stress gradient '$\sigma_R(x)$' is therefore obtained by adding the solid-liquid stress gradient '$\sigma_S(x)$' and the projection of the liquid-vapor stress gradient '$- \tau_{LV}(x)$' onto the derivative of the initial profile defined in equation (4), as follows :

$$\sigma_R(x) = \sigma_S(x) - \tau_{LV}(x) * Z_S'(x) \quad (11)$$

Where the derivative of the initial profile is given by : $Z_S'(x) = -\alpha\delta\, e^{-\alpha x}$
Recalling that the general form of the resulting stress is : '$\sigma_R(x) = \rho g\, B\, e^{-A x}$', we get :

$$\sigma_R(x) = \rho g\, B\, e^{-A x} = \rho g\, \delta\, e^{-\alpha x} + \rho g\, \lambda\, e^{-\varepsilon x} \,.\, \alpha\delta\, e^{-\alpha x} \tag{12}$$

Note that in our approach, since we know the meniscus profile '$Z_S'(x)$', it's easier to use the profile derivative than to measure an approximate angle using a visual method as is usually done with surface tension vectors in the Young-Dupré equation.

The gradients are related to the forces shown in **Figures 4a** and **4b** :

- The stress gradient '$\sigma_{S}(x)$' results from attractive forces causing vertical deformation and drawing the meniscus
- The stress gradient '$-\tau_{LV}(x)$' results from repulsive forces orthogonal to the curve generated by '$\sigma_{S}(x)$', tending to flatten it.

The meniscus profile is simply :

$$Z(x) = B\, e^{-A x} = \delta\, e^{-\alpha x} + \lambda\, e^{-\varepsilon x} \,.\, \alpha\, \delta\, e^{-\alpha x} \tag{13}$$

To plot the meniscus profile according to equation (13), we obviously need to know all the following parameters : 'B', 'A', 'δ', 'α', 'λ' and 'ε'.
Boundary conditions and the measurable value of the meniscus height 'B' can be used to calculate some parameters. For example, along the wall, at '$x = 0$' and with the measurement of 'B', the value of 'α' can be calculated as : $\alpha = (B - \delta)/\lambda\delta$. The other parameters will be determined later in the case of water, using measured values from the literature and through analysis of cases such as the capillary tube, the Wilhelmy plate (measurement of 'α' and 'δ') and the water drop.
As an example, **Figure 5** shows the water meniscus profile using the numerical values given below.

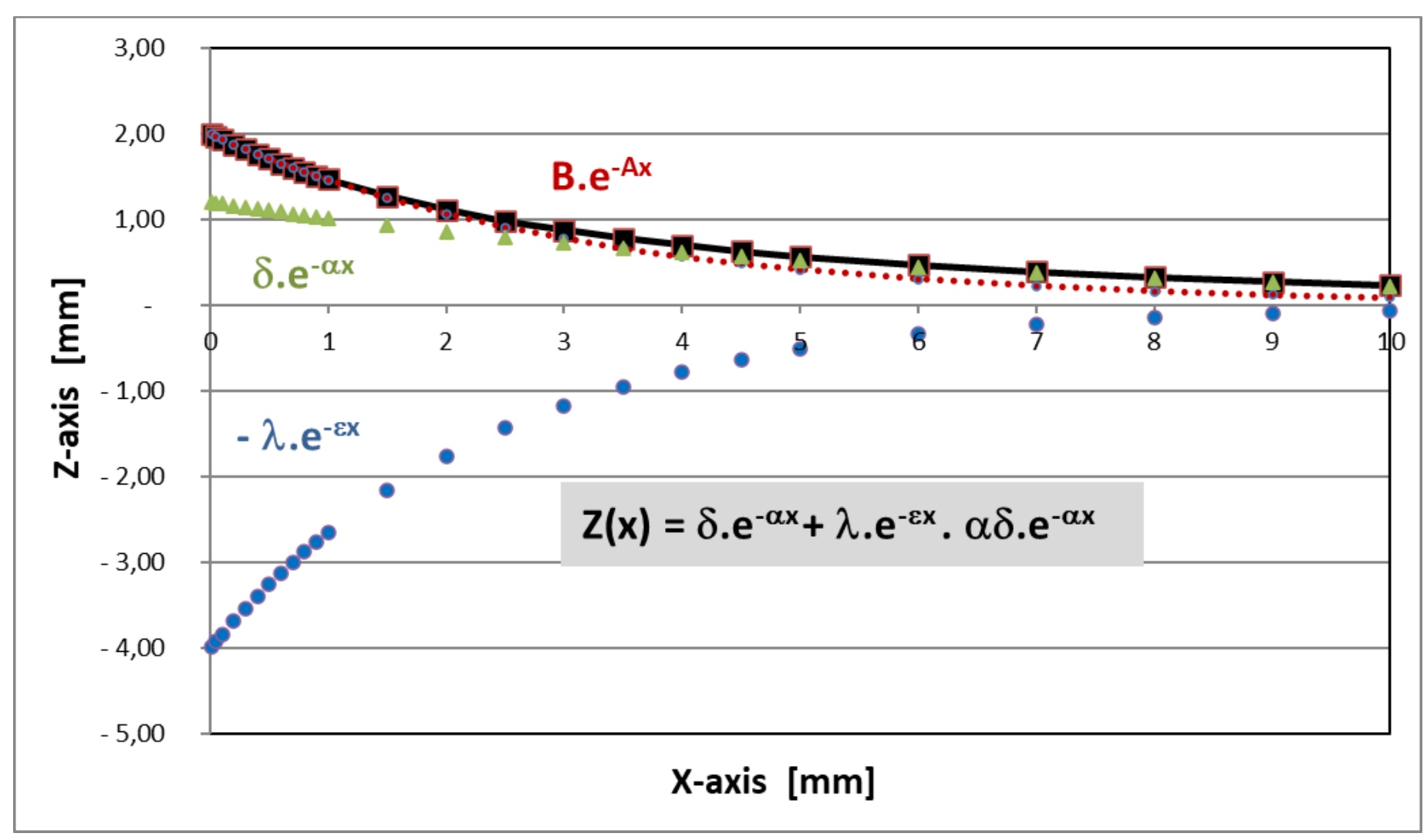

**Figure 5 – Calculated meniscus profiles water/glass with two gradients using following parameters : {B = 2.10$^{-3}$ [m] ; A = 316 [m$^{-1}$] ; δ = 1,2.10$^{-3}$ [m] ; α = 167 [m$^{-1}$] ; λ = 4.10$^{-3}$ [m] ; ε = 409 [m$^{-1}$]}**

**Figure 5** shows that the resulting profile '$Z(x) = B\, e^{-A\,x}$' from equation (4) is very close to '$\sigma_R(x)/\rho g$' of equation (12). That is why, in the following practical examples, we will rather use the resultant '$\sigma_R(x) = \rho\, g\, B\, e^{-A\,x}$' because it is easier to handle.

## 2.6 Meniscus equation with a theoretical solid-vapor gradient and hypothesis of a precursor film above the meniscus

As we saw above, due to the mechanical equivalence model nature, the gradient related to surface tension forces at solid interfaces '$\sigma_S(x)$' cannot be decomposed into solid-liquid '$\sigma_{SL}(x)$' and solid-vapor '$\sigma_{SV}(x)$' components.
Despite this, and even though it is physically incorrect, we will use equation (10) considering that '$\sigma_S(x)$' represents a hypothetical solid-liquid gradient '$\sigma_{SL}(x)$' and add a hypothetical solid-vapor stress gradient '$\sigma_{SV}(x)$' such as :

$$\sigma_{SV}(x) = \rho\, g\, \kappa\, e^{-\nu\, x} \tag{14}$$

where :

$\sigma_{SV}(x)$ : could be the solid-vapor interface stress gradient in [$Nm^{-2}$] or [$Jm^{-3}$]
$\kappa$ : is a parameter linked to the film height [m]
$\nu$ : is an attenuation constant [$m^{-1}$]

This mathematical trick will allow us to introduce the hypothesis of a "precursor film" above the meniscus. When analyzing the case of a hemispherical drop deposited on a solid surface, we will see that a number of observers have shown that there is a 'precursor film', a film less than 100 nanometers thick that extends beyond the base of the drop.

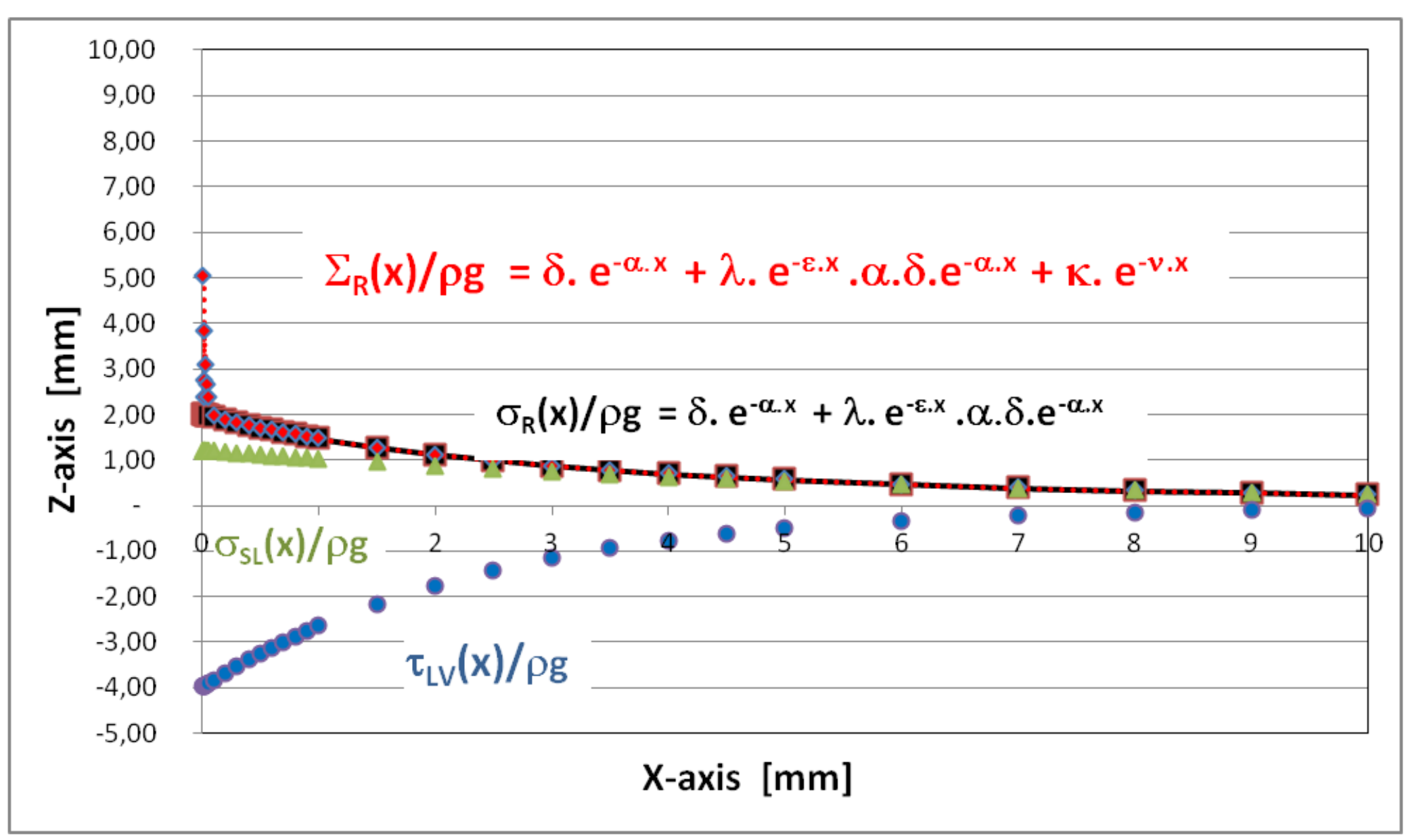

**Figure 5bis – Calculated meniscus profiles water/glass with a third theoretical gradient and with the following parameters : { B = 2.10$^{-3}$ [m] ; A = 316 [m$^{-1}$] ; δ = 1,2.10$^{-3}$ [m] ; α = 167 [m$^{-1}$] ; λ = 4.10$^{-3}$ [m] ; ε = 409 [m$^{-1}$]; κ = 5.10$^{-3}$ [m]; ν = 5.10$^{4}$ [m$^{-1}$] }**

In this article, we believe that there is also a precursor film above the meniscus.
Regarding the meniscus, for the moment, we will simplify the problem by considering only the two gradients at the liquid-solid and liquid-vapor interfaces to validate the shape of the meniscus.

This mathematical trick will allow us to introduce the hypothesis of a 'precursor film' above the meniscus. When analyzing the case of a hemispherical drop deposited on a solid surface, we will see that a number of observers have shown that there is a "precursor film," a film less than 100 nanometers thick that extends beyond the base of the drop.
In this article, we believe that there is also a precursor film above the meniscus corresponding to the solid-vapor constraint '$\sigma_{SV}(x)$'.
To model this film, we have modified equation (12) and **Figure 5** by adding a theoretical solid-vapor stress gradient and arbitrarily setting the values of 'κ' and 'ν', to plot the theoretical curve of **Figure 5bis**.

$$\sigma_R(x) = \rho\, g\, \kappa\, e^{-\upsilon\, x} + \rho\, g\, \delta\, e^{-\alpha\, x} + \rho\, g\, \lambda\, e^{-\varepsilon\, x} .\, \alpha\, \delta\, e^{-\alpha\, x} \qquad (15)$$

Equation (15) and **Figure 5bis** may not seem useful in the case of water, because if there is a precursor film above the meniscus, it is not visible due to its very low theoretical thickness and the transparency of water. Therefore, to check for the presence of a precursor film, we will instead examine the example of mercury meniscus which, due to its non-transparency, should reveal such a film.

### 2.7 Application of our approach to the case of the convex meniscus of mercury

To validate our approach, we plotted **Figure 6a** using equation (15) for mercury. We estimated the parameters from values provided in the literature for mercury, namely : $\gamma_{LV} \approx$ 0,500 [$Nm^{-1}$] ; $\gamma_{SL} \approx$ 0,400 [$Nm^{-1}$]) ; $\rho$ = 13,546 $10^3$ [$kg\ m^{-3}$], and set the values of 'κ' and 'ν' arbitrarily.
Since mercury repels the solid wall, the value of 'δ' and 'λ' are negative. These values are of course indicative, but the form of the equations is instructive.
This allows us to validate our hypothesis and verify in the photo in **Figure 6b** that a precursor film is visible in the case of mercury : the precursor film is transparent at the back due to its low thickness, but it is visible by reflection of light at the front of the tube due to the mirror effect.

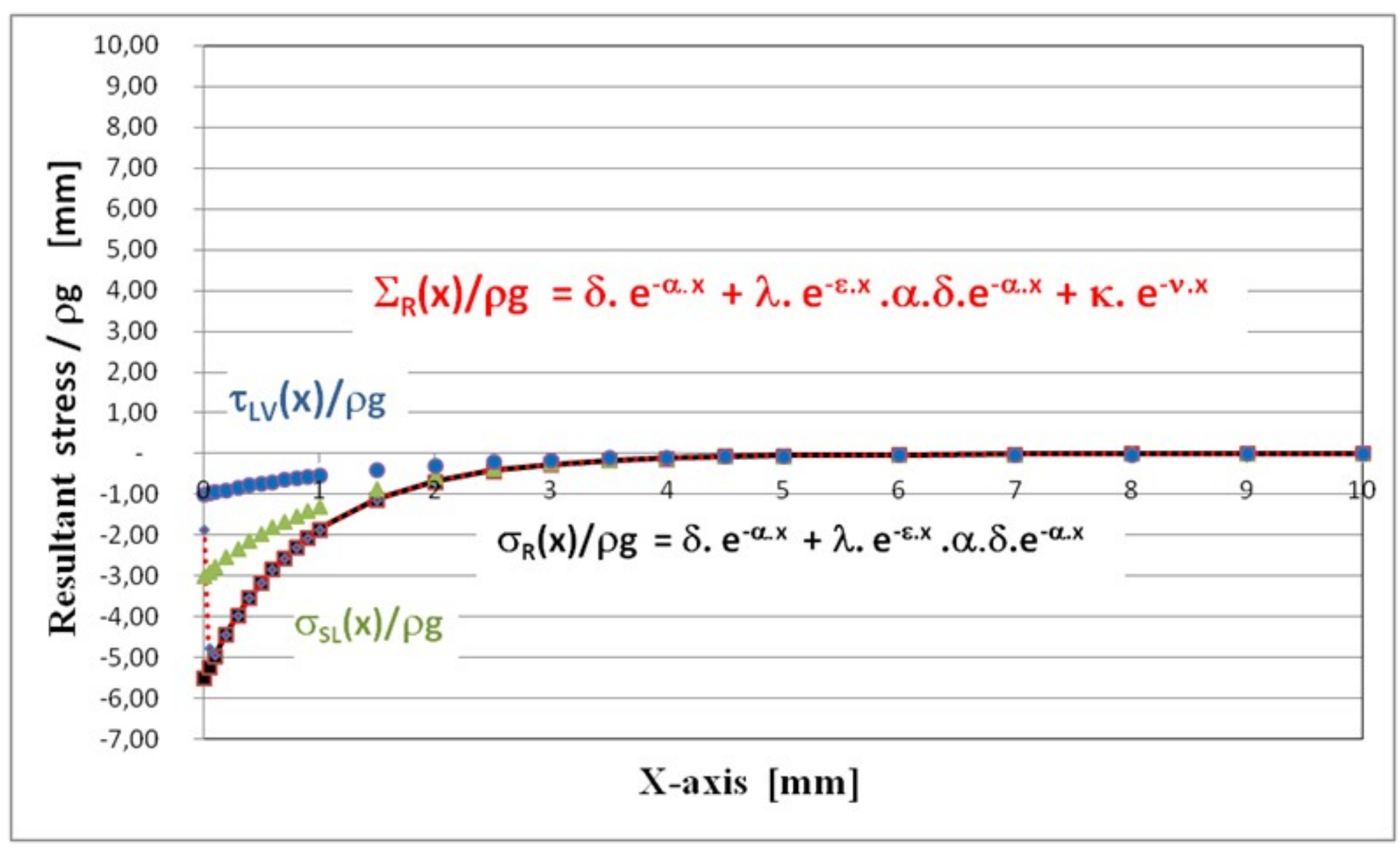

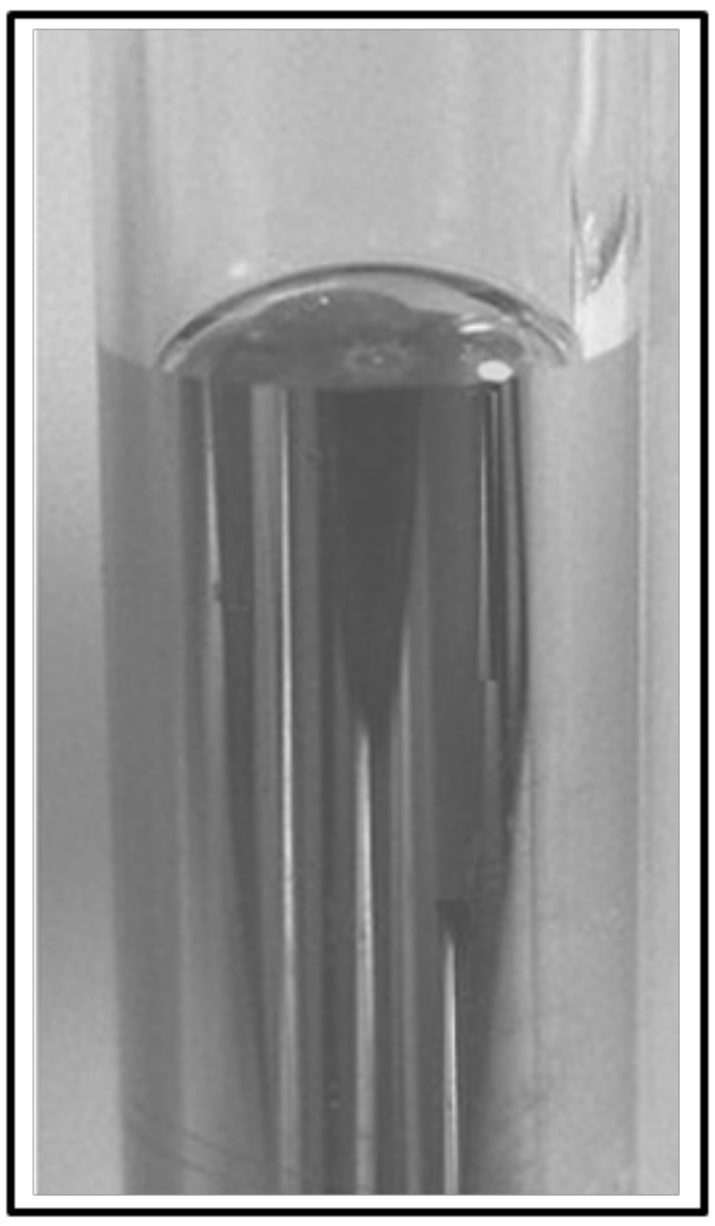

**Figure 6a – Calculated meniscus profiles for mercury using the following parameters :**
**{δ = – $3.10^{-3}$ [m]; α = 850 [$m^{-1}$]; λ = $1.10^{-3}$ [m]; ε = 650 [$m^{-1}$]; κ = $6.10^{-3}$ [m] ; ν = $5.10^4$ [$m^{-1}$]}**
**Figure 6b – Overview of the mercury meniscus in a glass tube showing the presence of a precursor film**

The photo of **Figure 6b** demonstrates the presence of a precursor film located above the meniscus.

We will return to this concept of precursor film later. In the meantime, we will apply our model to practical cases studied in the literature (Wilhelmy tensiometer, capillary tube, drops, etc.).

# III Application of our model to practical cases

As part of our thought experiment, we will use the above-defined equations and check their relevance in practical cases such as the Wilhelmy plate tensiometer and the capillary tube case (Jurin's law). Finally, we will calculate the pressure variation in a drop as in Laplace's law.

## 3.1 The Wilhelmy plate tensiometer

The Wilhelmy[13-15] plate tensiometer is a device for measuring the surface tension of a liquid at equilibrium. The apparatus uses a thin plate connected to a microbalance, as shown in **Figure 7**. The plate is perpendicular to the air-liquid interface and the force exerted on this plate is measured.
According to Wilhelmy, the force measured by the tensiometer '$F_{mes}$' is the vertical component of the surface tension force '$F_{TS} = \gamma L$', such as : $F_{TS} = F_{mes} / \cos(\theta)$.
The surface tension '$\gamma$' can be calculated as follows :

$$\gamma = \frac{F_{mes}}{L \cos(\theta)} \tag{16}$$

Where :

$\gamma$ : is the measured surface tension [N/m]
$F_{mes}$ : is the force measured by the microbalance [Kg]
L : is the plate perimeter (L= 2w+2d) [m]
w : is the plate length [m]
d : is the plate thickness [m]
$\theta$ : is the contact angle between the plate and the liquid

In the literature, the surface tension '$\gamma$' measured by this method is generally considered to be the liquid-vapor surface tension '$\gamma_{LV}$'.

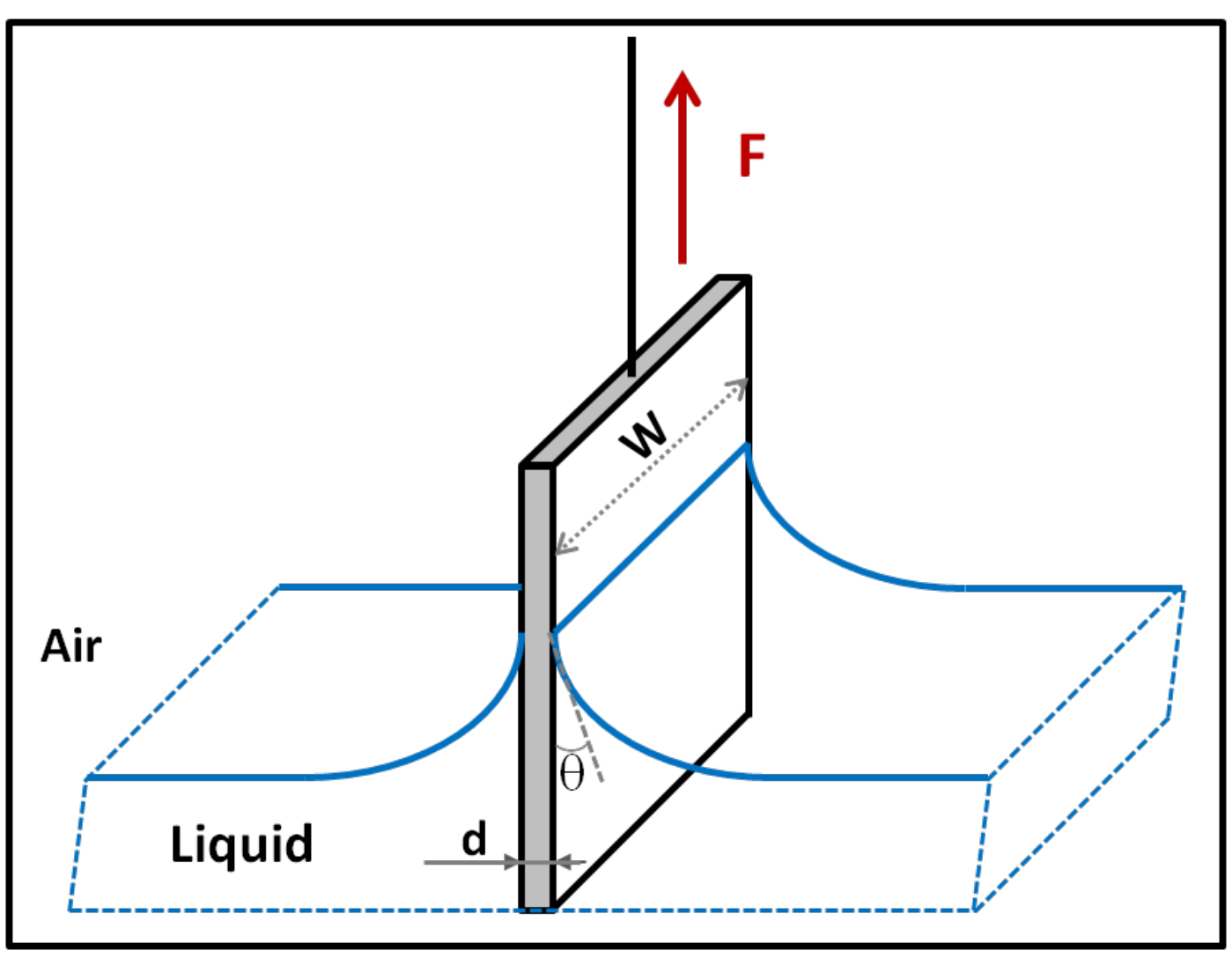

**Figure 7 - Wilhelmy plate : The capillary force is proportional to the plate perimeter and to the surface tension '$\gamma_{LV}$'**

### 3.2 Application of our approach to the Wilhelmy plate tensiometer

We can use the profile equation (13) to calculate the meniscus weight '$P_m$' as : $P_m = \rho\ g\ V_m$.
'$V_m$' is the meniscus volume. The volume is calculated by integrating the meniscus profile equation to infinity :

$$V_m = L\ B \int_0^{\infty} e^{-A\,x}\ dx, \text{ that is : } V_m = L\ \frac{B}{A}.$$

Since the measured force is equal to the meniscus weight, it is calculated as follows :

$$F_{mes} = \rho\ g\ L\ \frac{B}{A} \qquad (17)$$

Where :

$F_{mes}$ : is the force measured by the balance [Kg]
L : is the plate perimeter [m]
B : is the meniscus height at the wall [m]
A : is an attenuation constant to determine [$m^{-1}$]

Note that there is no '$\cos(\theta)$' factor in our equations because we consider that the force '$F_{mes}$' comes from the meniscus weight and that the meniscus is the resultant of the stress gradients defined in equation (12).
Using the value of '$F_{mes}$' measured by the tensiometer, the damping constant 'A' can then be calculated as follows :

$$A = \rho\ g\ L\ B\ /\ F_{mes} \qquad (18)$$

Equation (18) can be used to calculate 'A' knowing the value of '$\gamma_{LV}$' given in the literature and the meniscus height 'B' as : $A = \rho\ g\ B\ /\ (\gamma_{LV} \cos(\theta))$.
In the case of water, using the following parameters : $g = 9.81$[N $Kg^{-1}$] , $\rho = 10^3$ [Kg $m^{-3}$] and $B = 2.10^{-3}$ [m], the value of the surface tension given in the literature : $\gamma_{LV} \approx 72.\ 10^{-3}$ [$Nm^{-1}$] and with an angle '$\theta$' of about thirty degrees, we obtain : $A = 316$ [$m^{-1}$].
It should be noted that these are the values used in **Figure 5**.

Furthermore, if we compare Wilhelmy's surface tension equation (16) with our equation (17), we can define a term '$\Gamma$', which is equivalent to surface tension, such as :

$$\Gamma\ = \rho\ g\ B/A = \gamma_{LV} \cos(\theta) \qquad (19)$$

In the case of water, with the above parameters, we get : $\Gamma = \gamma_{LV} \cos(\theta) \approx 62.\ 10^{-3}$ [N $m^{-1}$].
The definition of the term '$\Gamma$' will be discussed later in section VI.

### 3.3 The capillary tube case (Jurin's law)

The phenomenon of capillary rise has been described by James Jurin[2] in 1718.
According to Jurin, the height of liquid in a capillary tube is inversely proportional to the radius of the tube. This phenomenon, illustrated in **Figure 8**, has been expressed mathematically using the hydrostatic law in the tube ($\Delta p = \rho\ g\ h$) and the Young-Laplace equation ($\Delta p = 2\ \gamma \cos(\theta)\ /\ r$), where '$\gamma$' has been replaced by '$\gamma \cos(\theta)$' in the Laplace's equation, assuming that the profile of the surface is in the form of an arc and that the radius of curvature at the air-liquid interface is equal to ' $r/ \cos(\theta)$'.
Thus, the Jurin' law is expressed as :

$$h = \frac{2\ \gamma_{LV}\ \cos\ (\theta)}{\rho\ g\ r} \tag{20}$$

Where :

h : is the height of fluid in the capillary tube [m]
$\gamma_{LV}$ : is the liquid-vapor surface tension [N $m^{-1}$]
$\theta$ : is the contact angle between plate and liquid
$\rho$ : is the density [Kg $m^{-3}$]
g : is the intensity of gravity [m $s^{-2}$] or [N/kg]
r : is the capillary tube radius [m]
r/ cos($\theta$) : is the radius of curvature at the liquid-vapor interface

Jurin's law is limited to the case of capillary tubes whose radius 'r' is significantly smaller than the length '$L_c$', called 'capillary length', according to inequality : $r < L_c$ where : $L_c = \sqrt{\frac{\gamma_{LV}}{\rho\ g}}$ .

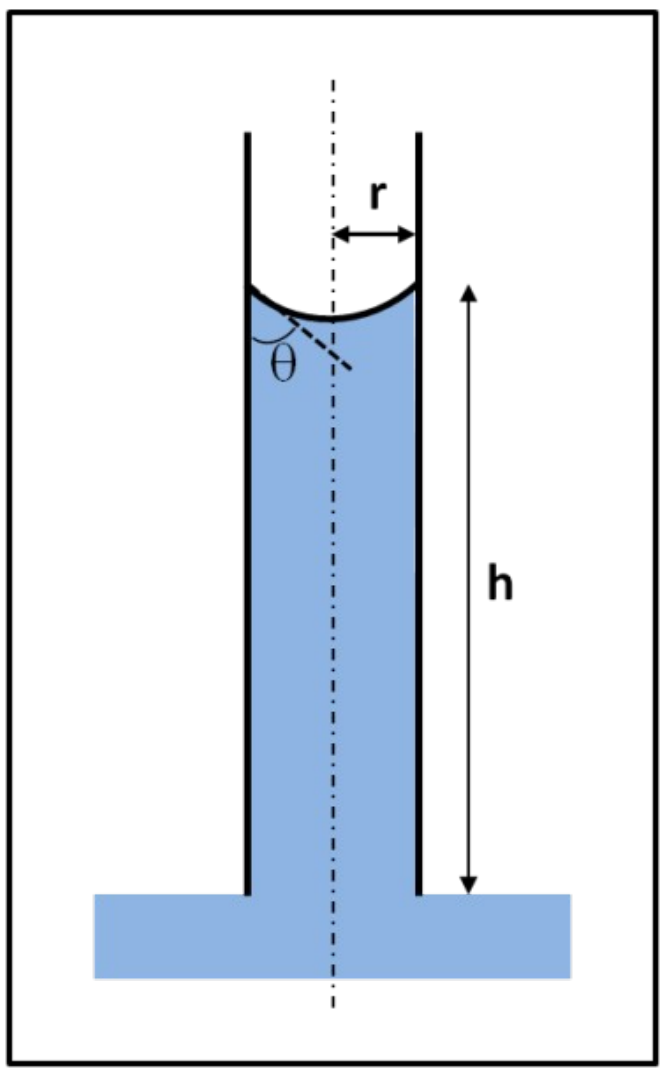

**Figure 8 – Rise of the liquid in a capillary tube**

### 3.4 Application of our approach to the capillary tube case

According to our approach, there is a balance of forces acting inside and outside the tube.
Inside the tube, the force '$F_i$' is calculated as the product of hydrostatic pressure ($\Delta p = \rho g h$) by the tube section ($\Pi r^2$) :

$$F_i = \rho g h . \Pi r^2 \tag{21}$$

Outside the tube, the surface tension force '$F_T$' acts on the perimeter of the tube ($2\Pi r$) as in the case of the Wilhelmy blade :

$$F_T = 2\Pi r\ \rho g\ (B/A) \tag{22}$$

The liquid height in the capillary tube 'h' can be calculated by balancing the two forces as :

$$h = 2B/Ar \tag{23}$$

Using the tension surface equivalent 'Γ' from equation (18), equation (23) can be compared to that of Jurin (20) as :

$$h = 2\,\Gamma\,/\,\rho g\,r \tag{24}$$

Where : $\Gamma = (\rho g\,B/A)$

In the case of water, using the following value of the meniscus height '$B = 2.10^{-3}$ [m]', the value of surface tension '$\gamma_{LV} \approx 72.\,10^{-3}$ [$Nm^{-1}$]', an angle 'θ' of about thirty degrees, corresponding to : $\Gamma \approx 62.\,10^{-3}$ [$N\,m^{-1}$]', the curves corresponding to (20) and (24) have been drawn in **Figure 9**.
We observe that the result is consistent with literature[16].

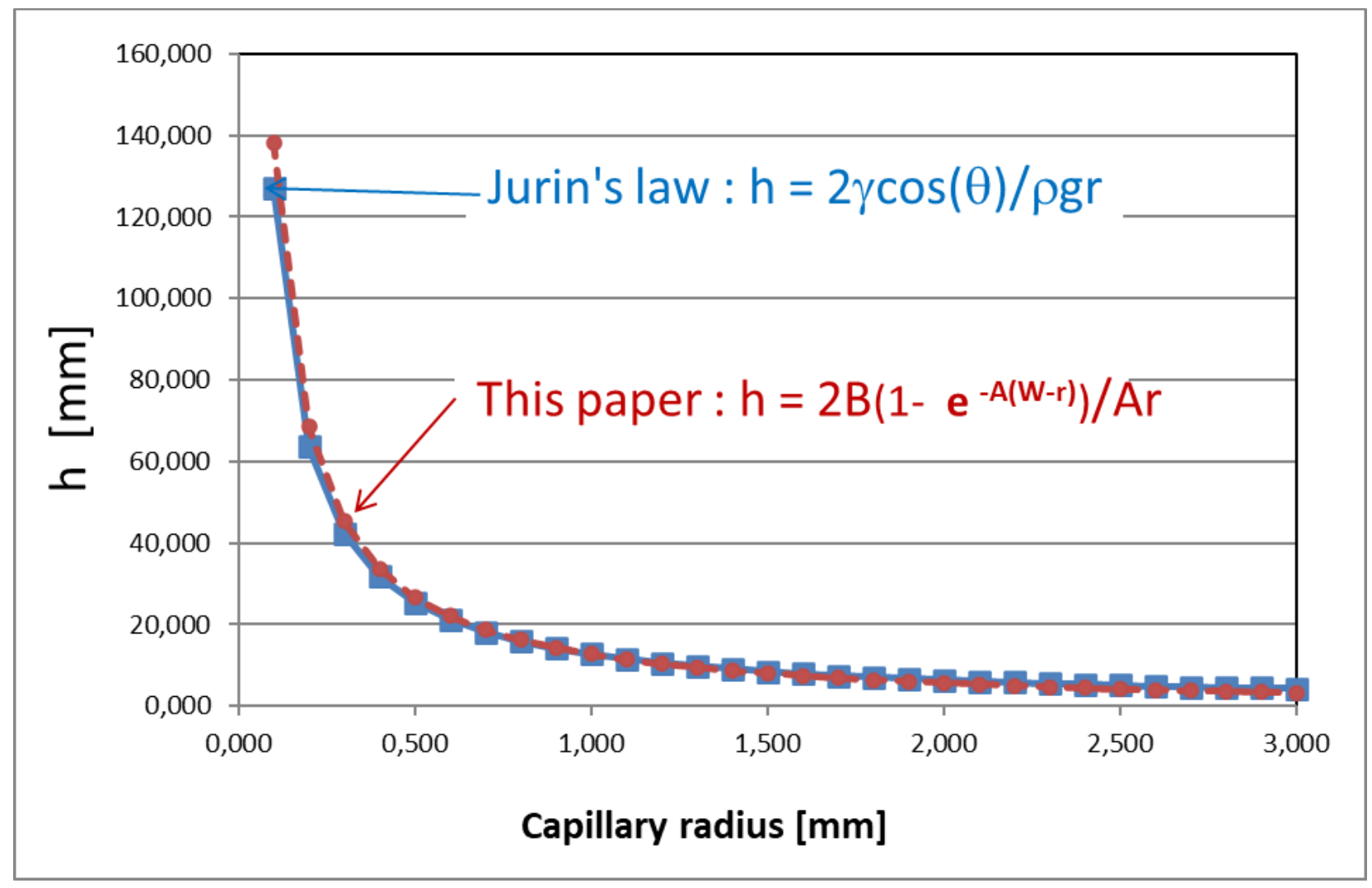

**Figure 9 - Compared curves of liquid rise in a capillary tube using following parameters: { $B = 2.10^{-3}$ [m]; A = 316 [m-1]; $\gamma_{LV} \approx 72.\,10^{-3}$ [$N\,m^{-1}$]; $\Gamma \approx 62.\,10^{-3}$ [$N\,m^{-1}$] }**

## IV The case of a spherical drop

Let us see how the notion of surface stress gradient can be used in the theoretical case of a spherical water drop.
According to the Laplace equation, the pressure difference 'ΔP' between the inside and outside of a spherical drop is calculated by balancing the resultant of the pressure forces acting on the equatorial surface '$\Delta F = \Pi R^2.\,\Delta P$' and the surface tension force exerted on its perimeter '$2\Pi R.\,\gamma_{LV}$'.

The result is the well-known Laplace equation : $\Delta P = \dfrac{2\,\gamma_{LV}}{R}$ .

This equation has its limits :

- The equation has an obvious infinity problem when the radius 'R' tends to zero.
- The equation is defined as being valid for : R << Lc where : $L_c = \sqrt{\dfrac{\gamma_{LV}}{\rho\ g}}$ .

Note that in the case of water, with the values indicated above, i.e. : $\rho = 10^3$ [$Kg/m^3$], g = 9,81 [N/kg]) and $\gamma_{water} \approx 72.\,10^{-3}$ [N/m], we get : $L_c = 2,7$ mm.

In the literature, the value '$L_c$' is often considered as the critical radius of the drop. We will see below that our definition of the critical radius '$R_c$' is significantly different.

**4.1 Application of our approach to the spherical drop case**

In the theoretical case of an isolated spherical drop, whereas there is no contact with any solid surface, the only superficial forces are those located at the liquid-vapor interface, forces that produce radial compression. As the liquid-vapor interface stress gradient '$\tau_{LV}$' defined in equation (10) is radial, it can be written along the 'z' axis (from the surface where : z=0) as in **Figure 10a** :

$$\tau_{LV}(z) = - \rho\, g\, \lambda\, e^{-\varepsilon z} \quad (25)$$

where :

$\tau_{LV}(z)$ : is the liquid-vapor interface stress gradient in [$Nm^{-2}$] or [$Jm^{-3}$]
$\lambda$ : is a characteristic length of curvature in [m]
$\varepsilon$ : is an attenuation constant [$m^{-1}$]
$W_g$ : is an attenuation length [m]

As we did in the case of the meniscus with 'Wm', we can define here a theoretical attenuation length '$W_g$', corresponding to the theoretical maximum effective distance of the liquid-vapor interface forces.

Boundary conditions :
According to the literature, the cohesion of a drop depends on its size. Namely, when the radius of the drop reaches a critical radius '$R_c$', gravitational forces become predominant and break up the drop, as shown in **Figure 10a**.
On the other hand, we can no longer assume, as in the case of the meniscus, that the surface of the liquid is significantly larger than the distance of action of the surface tension forces, and then we can no longer assume that the set of forces is in equilibrium.
This means, among other things, that the liquid-vapor interface stress gradient '$\tau_{LV}(z)$' is maximum at the drop surface (when : z = 0), but it is not zero in the center of the drop (when : z = R) as represented in **Figure 10b**, because the critical radius is smaller than the attenuation length '$W_g$'.

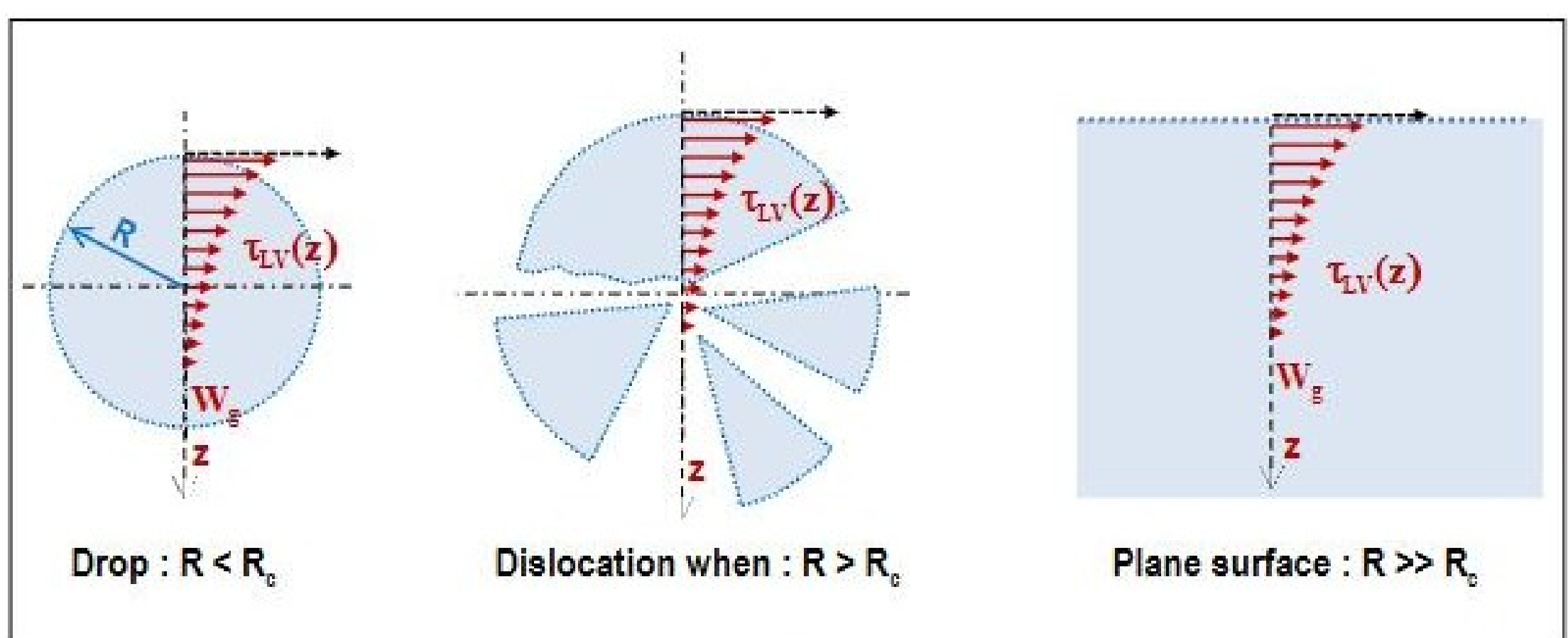

**Figure 10a – Liquid-vapor interface stress gradient effect at the liquid-vapor interface of a drop and at the surface of a liquid**

**Figure 10a** helps to explain the action of the liquid-vapor interface forces :

- A spherical drop only forms if its radius is less than the critical radius '$R_c$'.

- When the radius of the drop exceeds the critical radius '$R_c$', gravity forces are predominant and dislocate the drop.
- For a large area of water, the liquid-vapor interface forces cannot curve the surface because it has an almost infinite dimension.
- When the drop radius is equal to the critical radius '$R_c$', the forces of gravity and the liquid-vapor interface forces are in equilibrium. This will enable us to derive the equilibrium equation between the drop weight and the integral of the liquid-vapor interaction stress gradient.

The stress gradient at the liquid-vapor interface '$\tau_{LV}(x)$' can be plotted as in **Figure 10b**, this time with the drop center placed at the origin of an 'x0z' coordinate system. This formalism will be used later to analyze other shapes of drops such as hemispheres or semi-ellipsoids deposited on a horizontal surface.
Then we can rewrite '$\tau_{LV}(x)$' as :

$$\tau_{LV}(R-x) = - \rho\, g\, \lambda\, e^{-\varepsilon\,(R-x)} \qquad (26)$$

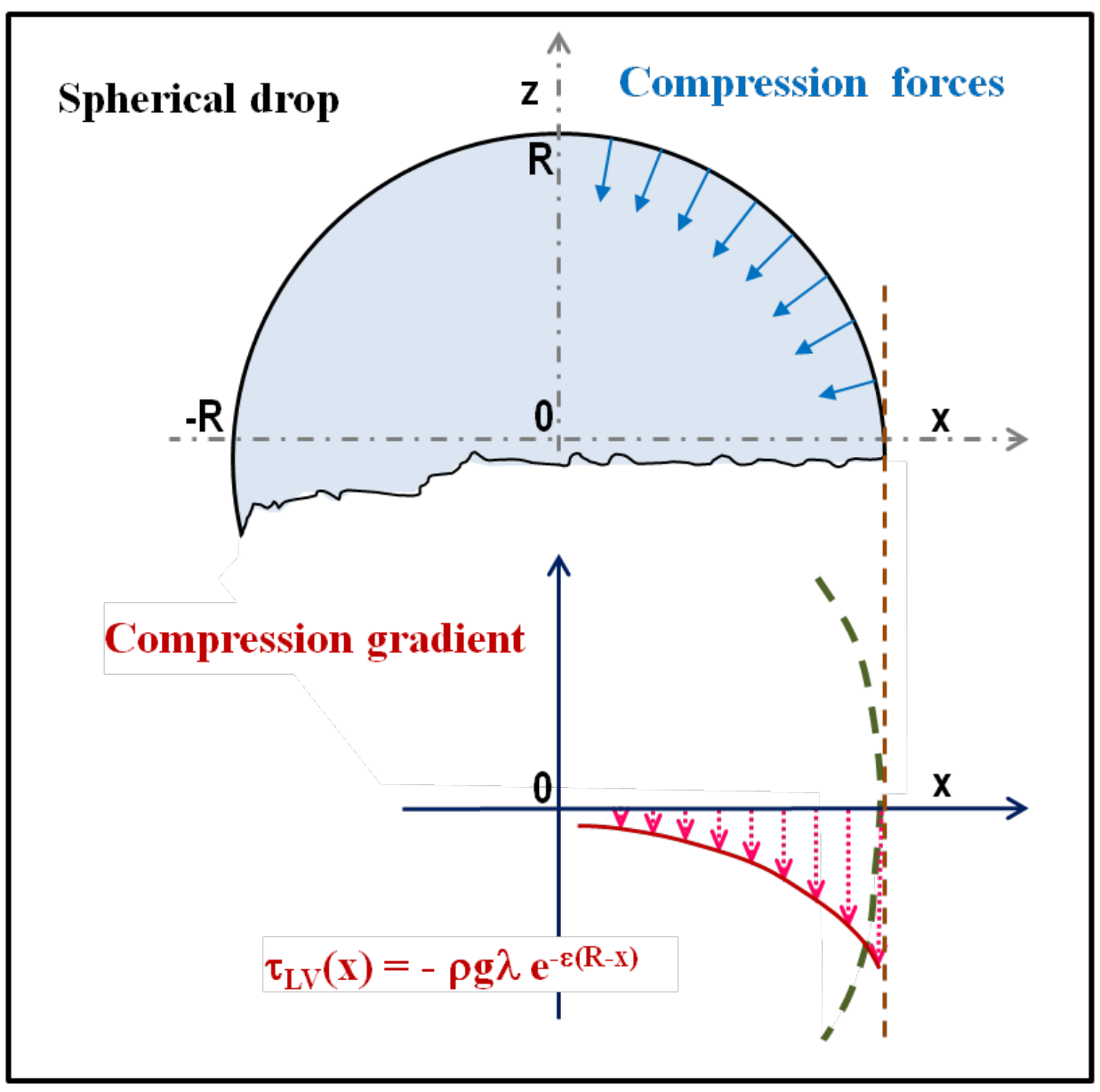

**Figure 10b – Compression gradient '$\tau_{LV}(x)$' in a spherical drop of water due to compression forces. The gradient is maximum at the surface and decreases toward the center**

The gradient is maximum at the drop surface and decreases towards the center :

- At the surface of the drop, we have : $\tau_{LV}(x=R) = - \rho\, g\, \lambda$
- At the center of the drop, the gradient value is : $\tau_{LV}(x=0) = - \rho\, g\, \lambda\, e^{-\varepsilon R}$ .

As mentioned above, the gradient value at the center of the drop is not zero because the radius is less than the attenuation length ($R_c < W_g$).
According to our equations, the liquid-vapor stress gradient should be zero at attenuation distance '$W_g$'. However, as mentioned above, it cannot be observed in the gravitational field, because the force of gravity will dislocate the drop when its radius exceeds the critical radius '$R_c$'. We will see later that this is no longer the case in weightlessness.
When the drop radius is equal to the critical radius '$R_c$', we can write the equilibrium equation between the critical weight '$P_c = \rho\, g\, (4/3)\, \Pi\, R_c^3$' and the sum of the forces exerted by the gradient on the drop perimeter '$2\pi R_c$' as :

$$\Sigma\ \text{Forces} = - 2\, \Pi\, R_c \int_0^{Rc} \rho\ g\ \lambda\ e^{-\varepsilon\,(Rc-x)}\, dx \qquad (27)$$

Which leads to equation (29) :

$$\frac{2}{3}\, R_c^2 = (\lambda/\varepsilon)\,(1 - e^{-\varepsilon R_c}) \qquad (28)$$

This equation can be used to calculate the characteristic length '$\lambda$' as : $\lambda = \frac{2}{3} R_c^2\, \varepsilon / (1 - e^{-\varepsilon R_c})$.

For critical radius values such as : $R_c = 3$ to 4 mm[14] and : $\varepsilon = 409$ [$m^{-1}$], we get a characteristic length value '$\lambda$' such as : $\lambda$= 3,5 to 5 mm.

NB : In the following examples, we will use : $\lambda = 4$ mm and $R_c = 3{,}3$ mm.

As can be seen in **Figure 10a**, when the drop radius is less than the critical radius '$R_c$', equation (28) becomes an inequation such as :

$$\frac{2}{3}\, R^2 < \lambda/\varepsilon\,(1 - e^{-\varepsilon R}) \qquad (29)$$

We can check that the drop is in compression as long as : $R \leq R_c$.

We can now calculate the pressure variation in the drop as in Laplace's equation.

**4.2 Calculation of the pressure difference in the drop and Laplace's law**

As for the Laplace equation, we obtain the drop pressure variation '$\Delta P$' in the median plane of the sphere by calculating the sum of the forces exerted on its perimeter '$2\Pi R$' and dividing it by the area of the median plane '$\Pi R^2$' as :

$$\Delta P = \int \text{Forces} / \Pi R^2$$

The sum of the forces is obtained by integrating the liquid-vapor interface stress gradient '$\tau_{LV}(r)$' along the perimeter '$2\Pi R$' using equation (28) as :

$$\int \text{Forces} = -\,2\Pi R\,\rho\, g\,\lambda \int_0^R e^{-\varepsilon\,(r-x)}\,dx \qquad (30)$$

Which leads to :

$$\Delta P = (2/R)\,\rho\, g\, \frac{\lambda}{\varepsilon}\,(1 - e^{-\varepsilon R}) \qquad (31)$$

That can be rewritten as :

$$\Delta P = 2\,\rho\, g\,\lambda\,\{(1 - e^{-\varepsilon R})/(\varepsilon R)\} \qquad (32)$$

The expression between the brackets '$\{(1- e^{-\varepsilon R})/(\varepsilon R)\}$' is interesting because it tends to unity when '$R$' tends to zero, which means that '$\Delta P$' tends to '$2\,\rho\, g\,\lambda$'.

Furthermore, using the term ‘Γ’ defined above, equation (32) can be rewritten by analogy with Laplace equation:

$$\Delta P = \frac{2\ \Gamma}{R} \tag{33}$$

Where ‘Γ’ is equivalent to surface tension. Here, its value is : $\Gamma = \rho\ g\ (\lambda/\varepsilon)\ (1 - e^{-\varepsilon R})$ [$Nm^{-1}$].
NB : The term ‘Γ’ will be discussed in more detail in Section VI.

Let us compare equations (32) and (1) as a function of ‘R’ in **Figure 11** :
When ‘R’ tends to zero, the pressure in equation (1) tends to infinity, whereas in equation (32), pressure tends to a finite value : $\Delta P(R=0) = 2\ \rho\ g\ \lambda$.
It should be noted that the theoretical pressure difference in a 1 micron diameter drop should be approximately 2,8 bars according to equation (1), whereas it should be only of $87.10^{-5}$ bars according to equation (32), a value that seems more reasonable to us, as long as the rain microdrops do not explode when they fall to the ground.

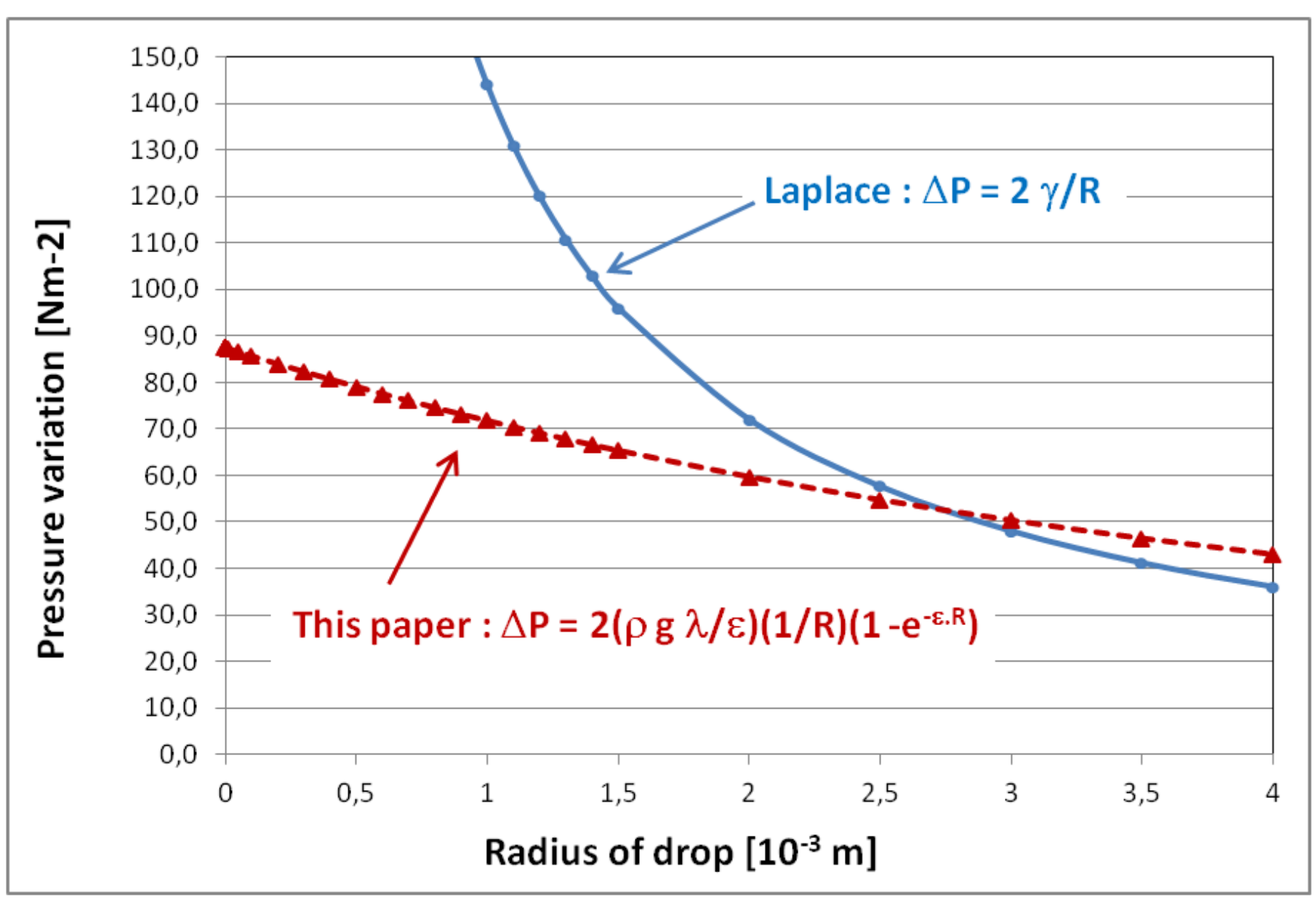

**Figure 10 – Equation (33) of the pressure variation in a drop: comparison with the Young-Laplace equation (1) {using following parameters : $\lambda = 4.10^{-3}$ [m]; $\varepsilon = 409$ [$m^{-1}$]}**

### 4.3 Maximum size of a drop and critical radius

According to most authors, the critical radius ‘$R_c$’ for water is about 3 mm, although, according to H. R. Pruppacher and J. D. Klett[17], water drops 8 mm in diameter ($R_c$ of about 4 mm) have been observed and measured.
In this paper, we have arbitrarily fixed the value of the characteristic length of curvature as : λ = 4 mm (in **Figures 5, 5bis, 11** and following). This corresponds to a critical radius ‘$R_c$’ of approximately 3,3 mm.

It should be noted that we have little information on the critical radius and that it has not really been the focus of research to date, probably because it was useless in previous equations (laws of Young-Laplace and Young-Dupré). In our approach, however, the measurement of '$R_c$' is much more important because it enables us to calculate the parameter ‘λ’.
Regarding the theoretical attenuation length '$W_g$', which cannot be observed on Earth due to gravity, it will be discussed in the following section.

## 4.4 Maximum size of a drop of water in weightlessness.

In Earth's gravitational field, we know that weight limits the size of water drops because they break apart when their radius reaches the critical radius '$R_c$'.
However, in zero gravity, our equations suggest that we should observe water drops with a much larger radius.
This is exactly what the cosmonauts observed during their experiments in weightlessness, when handling water drops more than ten centimeters in diameter[18-20].
Let us remember that in equations (5-7), the surface energy per unit volume 'u(x)' has been balanced with the potential gravity energy per unit volume '$E_{pp}(x)$' and a constant '$u_0$' has been calculated as a function of the intensity of gravity.
In zero gravity we should theoretically restart from the expression of the surface tension energy '$u(x) = u_0 e^{-Ax}$'.
Nevertheless, to simplify matters, we will use the expression (33) to calculate the pressure difference in the water drop and consider the 'Γ' parameter as independent of gravity. Alternatively, we can use the value of 'Γ' provided by equation (32) and calculate the pressure variation in the drop.
We can then plot the theoretical pressure variation in the drop above '$R_c$' as in **Figure 12** and extrapolate a hypothetical value of '$W_g$', which should be of the order of a dozen centimeters.
According to equation (33), the drop should theoretically remain in compression, even if the pressure variation is very small.

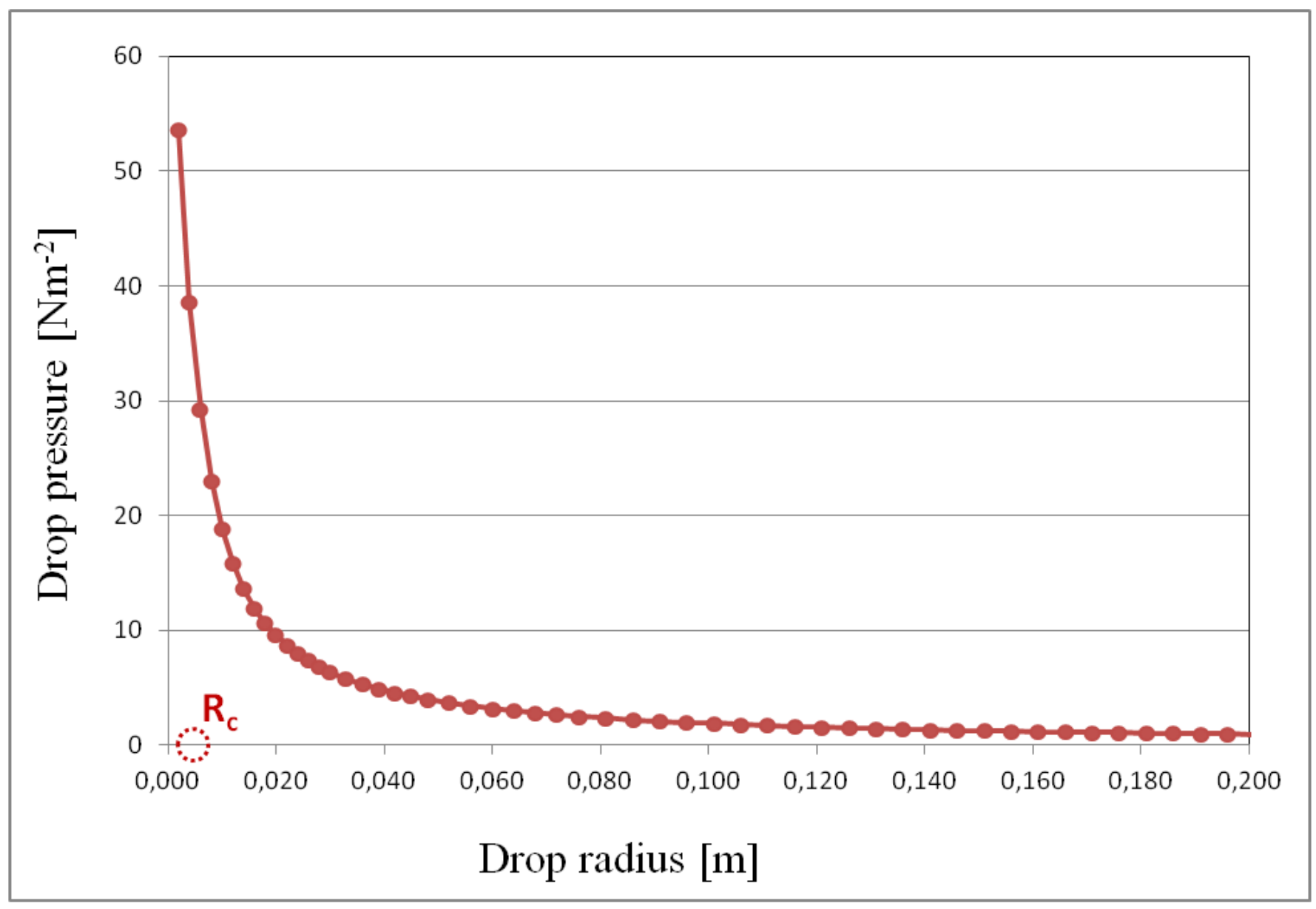

**Figure 12 – Theoretical water drop pressure variation in zero gravity**

Note that the calculated pressure variation remains very low compared to the pressure of human breath, around 0,1 bars or $10^4$ [$Nm^{-2}$]. This means that a drop in weightlessness should be easily deformed or dissociated with the slightest breath.
Although there is no information yet on the measurement of the attenuation length, the height of the meniscus or the maximum radius of curvature in microgravity, our hypothesis remains plausible and is consistent with the observation by cosmonauts of water drops with a diameter of more than a dozen centimetres[18-20] in zero gravity.

# V The case of a drop on a solid surface

## 5.1 The Young-Dupré equation

In the theoretical case of a drop deposited on a solid surface, the Young-Dupré[5] equation is usually used. According to this law and as described in **Figure 1**, the following three projected surface tensions are supposed be in equilibrium, as in equation (2) : $\gamma_{LV} \cos(\theta) = \gamma_{SV} - \gamma_{SL}$.
It should be noted that even if the three stresses projected onto the plane are in equilibrium, the vector sum of these three stresses includes a component perpendicular to the plane of the solid wall. This force is theoretically offset by the traction exerted by the solid on the line of contact.
According to the Young-Dupré equation, given that the surface tensions '$\gamma_{LV}$', '$\gamma_{SL}$' and '$\gamma_{SV}$' are fixed parameters depending on the type of solid, liquid, and vapor, there should be only one value of the contact angle '$\theta$'.
However, if we observe a series of water drops of different sizes deposited on the same surface, as in **Figure 13**, we can see that the drops vary in shape.
More precisely, we can observe that the small drops seem to have a shape ranging from a pseudo-sphere to a hemisphere, while the large drops tend to have a semi-ellipsoidal shape.
To explain the variations in contact angle, even though it is the same liquid and the same substrate, authors generally assume that if the contact angle varies from one drop to another, it is due to surface roughness and impurities causing a deviation from the contact angle predicted by the Young-Dupré equation.
Some authors do not agree with this statement. In his study of advancing or receding drops, Rafael Tadmor[21] recognizes that the contact angle depends on the volume, while Lasse Makkonen[22], on the other hand, considers that the contact angle does not depend on roughness or impurities, but that when a drop slides, the solid-vapor interface disappears and a solid-liquid interface is formed at the advancing contact line. Accordingly, the solid-liquid interface disappears, and a solid-vapor surface is formed at the receding contact line.

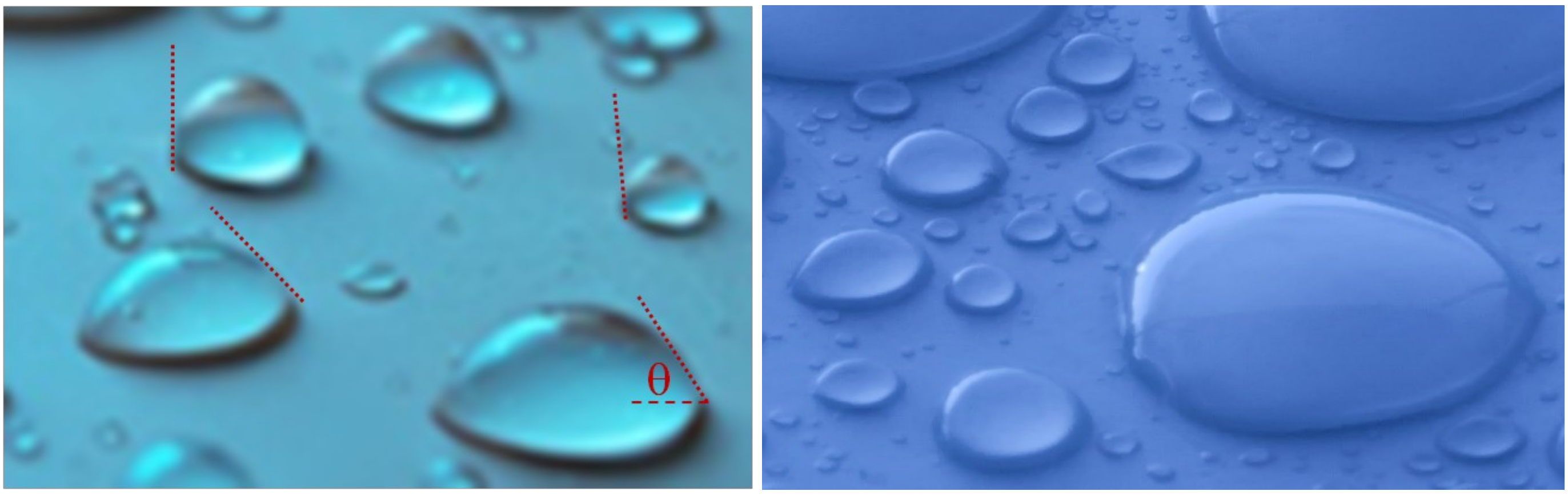

**Figure 13 - Photos of water drops of different sizes and shapes**

In this paper, rather than trying to interpret the Young-Dupré equation, we will use our equations to model existing drop shapes and validate them.

## 5.2 Preamble to the application of our approach to the case of drops on a solid surface

**Geometric conditions for modeling**
To model the behavior of real drops, we need to define drops with perfect symmetrical and regular geometric shapes because these shapes can be modelled in geometry.
In this paper, we have chosen to study theoretically 'hemispherical', 'semi-ellipsoids' and 'pseudo-spheres' drop like those usually depicted in literature.

**Boundary conditions :**
In the case of the spherical drop analyzed in section 4.1, we have seen that the surface area of the drop cannot be considered large in relation to the dimensions of the drop, so the forces can no longer be considered in equilibrium.
Similarly, in the case of a drop deposited on a solid surface, neither the liquid surface nor the solid contact surface can be considered large, and we cannot consider that the forces are in equilibrium.
In fact, a balance of forces is achieved only when the radius is equal to the critical radius, i.e., when the weight of the drop is equal to the sum of the surface tension forces. As in the case of the spherical drop, the drop is always under compression, and it is this compression that keeps it in shape.
In the case of the spherical drop, we have seen that when the radius is greater than the critical radius, the drop breaks up. Differently, in the case of a hemispherical drop deposited on a solid surface, when the radius is greater than the critical radius, the drop is changing its shape.
This means that we must write here not equations, but inequations. Moreover, in the case of a hemispherical drop, we must consider that the equilibrium is reached only when the drop radius reaches its critical value, noted here as '$R_{ch}$'.

**Comment on the value of '$R_{ch}$'**:
The literature does not provide information on the critical radius value of a hemispherical drop, probably because authors usually use the Young-Dupré equation. In contrast, in our approach, the critical radius value is important because it is related to the parameters of the solid interface gradient.
It should also be noted that, in the case of a hemispherical drop, the critical radius value is not a limit because, when the radius is greater than '$R_{ch}$', gravity does not break up the drop as it does with a sphere, but rather causes it to transform into a larger non-hemispherical drop.

**Quid of precursor film at drop edge :**
We know that a number of experimenters have observed a deformation of the foot drop, due to the 'precursor film' phenomenon[23-25]. The 'foot' of the drop will not be considered in this article and will be analyzed at a later date.

### 5.3 Applying our approach to the case of a hemispherical drop deposited on a solid surface

Let us start by applying our approach to the theoretical case of a perfectly hemispherical drop.
Two types of interaction forces are considered in **Figure 14** : the liquid-vapor interaction forces at the drop surface, and the solid (solid-liquid and solid-vapor) interaction forces at the drop-solid contact interface.
The resultant of superficial tension forces is opposed to gravity forces :

- As long as the radius is smaller than '$R_{ch}$', superficial tension forces are greater than gravity forces.
- When the radius is equal to the critical radius, there is equilibrium between the forces.
- When the radius is greater than the critical radius, gravity transforms the hemispherical drop into a larger non-hemispherical drop, e.g., a hemi-ellipsoid, in order to achieve a new equilibrium.

In Young-Dupré's equation, the solid-vapor surface tension tensor '$\gamma_{SV}$' tends to spread the contact line of the drop, while the projection of the liquid-vapor surface tension tensor '$\gamma_{LV} . \cos(\theta)$' and the solid-liquid surface tension tensor '$\gamma_{SL}$' contract it.
According to our approach, we consider that the resulting solid interface forces (related to the gradient '$\sigma_S$') and the gravitational forces tend to spread the drop, while liquid-vapor interface forces (related to the gradient '$\tau_{LV}$') contract it radially and maintain its cohesion.
Apart from taking gravity into account, there is no fundamental difference between the two views.
Indeed, we have seen that in the mechanical equivalence model, the solid-liquid component '$\sigma_{SL}(x)$' and the solid-vapor component '$\sigma_{SV}(x)$' cannot be distinguished from the resulting gradient '$\sigma_S(x)$', and these components could be in the same direction as the vectors '$\gamma_{SL}$' and '$\gamma_{SV}$'.

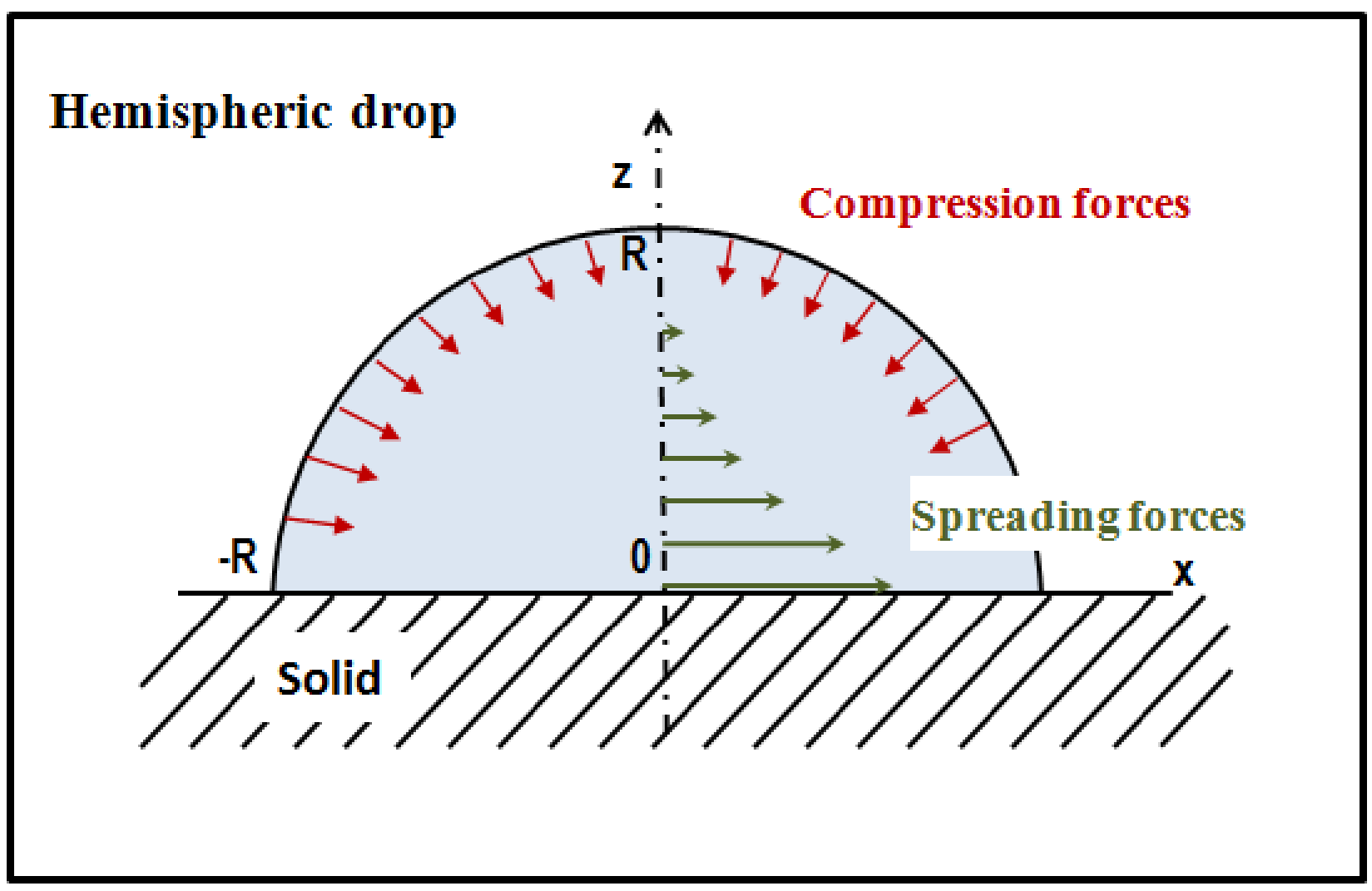

**Figure 14- Hemispheric drop on a solid surface. Drawing of compressive forces at the liquid-vapor interface and resulting spreading forces at solid interface**

Before calculating the components of these forces, we will begin by analyzing the stress gradients.

### 5.3.1 The liquid-vapor interface stress gradient

The liquid-vapor interface stress gradient '$\tau_{LV}(x)$' is radial. It acts perpendicular to the surface. As in the spherical drop case, it defines the drop shape. Since it is radial, it can be written equivalently in terms of 'x' or 'z', as below :

$$\tau_{LV}(z) = -\rho\, g\, \lambda\, e^{-\varepsilon z} \tag{34}$$

It is noted negatively because the drop is in compression.
Horizontally, as shown in **Figure 15a**, the liquid-vapor interface forces bring the drop edges towards the center. They are opposed to the resulting solid interface forces and the gravitational forces which tend to spread the drop.

**Boundary values**
The horizontal component of the gradient '$\tau_{LV}(x)$' is maximum at the edge of the drop :
$\tau_{LV}(z = 0\,;\, x = R) = \rho\, g\, \lambda$.
At the center of the drop, its value is minimal but not zero : $\tau_{LV}(z = R\,;\, x = 0) = \rho\, g\, \lambda\, e^{-\varepsilon R}$ .

### 5.3.2 The resulting solid interface stress gradient

The resulting solid stress gradient of a hemispherical drop deposited on a solid surface is different from that observed in the meniscus case.
Indeed, in the meniscus case, the stress gradient '$\sigma_S(x)$' represents a gradient of vertical deformation forces parallel to the wall (along the 'z' axis). It is opposed to gravity and decreases (along the 'x' axis) from the wall to the 'bulk'.
In the case of a hemispherical drop, the solid stress gradient corresponds this time to horizontal deformation forces parallel to the drop support plane (along the 'x' axis) and varies along the z-axis..
The resulting solid stress gradient can be written as in equation (35), where we kept the attenuation constant '$\alpha$' of equation (9) and where the maximum stress parameter is labeled '$\beta$', assuming that its value should be different from the meniscus value '$\delta$'.

The gradient '$\sigma_S(z)$' is limited by the drop edges and, like gravitational forces, it theoretically tends to distort the hemisphere created by the surface tension liquid-vapor forces.
As illustrated in **Figure 15a**, these deformation forces are parallel to the support and the gradient can be written as :

$$\sigma_S(z) = \rho\, g\, \beta\, e^{-\alpha z} \tag{35}$$

where :

$\sigma_S(z)$ : is the resulting solid interface stress gradient [$Nm^{-2}$] or [$Jm^{-3}$]
x : is the axis parallel to the support [m]
z : is the axis perpendicular to the support [m]
$\beta$ : is the maximum stress parameter [m]
$\alpha$ : is an attenuation constant [$m^{-1}$]
$W_h$ : is the exponential attenuation length [m]

**Limit values**
Horizontal solid forces are maximum at the solid surface level (z=0) and minimum at the top of the drop.
The gradient '$\sigma_S(z)$' is maximum at the solid surface level :

$\sigma_S(z = 0 ; x = R) = \rho\, g\, \beta$.

However, at the drop top, the solid-liquid stress gradient is minimal but not zero :

$\sigma_S(z = R ; x = 0) = \rho\, g\, \beta\, (1 - e^{-\alpha R})$.

As indicated above, liquid-vapor forces oppose both solid interface forces and the spreading imposed by gravity. They are large enough to force the drop into a hemispherical shape, whereas solid-liquid forces and gravity forces tend to spread it out. We assume also that the solid-liquid spreading forces are weaker than those observed in the meniscus case. We will see later that the solid stress parameter of the hemispherical drop '$\beta$' is indeed weaker than the stress parameter '$\delta$' of the meniscus.

### 5.3.3 Hemispherical drop stress gradients

As represented in **Figure 15b**, the horizontal stress resultant in hemispherical drop can be written as :

$$\Sigma^H_{Res}(z) = -\,\tau_{LV}(z) + \sigma_S(z) + \rho g\, z(x).z'(x) \tag{36}$$

Where the gravity gradient component is obtained by projection onto the drop profile and where :

- $\tau_{LV}(z)$ : is the liquid-vapor interface stress gradient, negative because in compression
    - It is applied radially around the perimeter '$2\Pi R$' and on an arc of height 'R' in the case of hemisphere
    - In absolute value, it is maximum at : z = 0 and minimum at : z = R
- $\sigma_S(z)$ : is the solid-liquid interface stress gradient
    - It is applied around the perimeter '$2\Pi R$' and on an arc of height 'R'
    - It is maximum at : z = 0 and minimum at : z = R
- $E_{pp}$ : is the pressure gradient (or potential gravity energy per unit volume)
    - It is applied vertically on the support surface '$\Pi R^2$' and its horizontal component is obtained by projection on the drop profile as : $E_{pp} = \rho g\, z(x).\, z'(x)$
    - The pressure gradient is maximum at : x = 0 (z = R) and minimum at : x = R (z = 0)
    - Its integration calculates the weight as : $P_{gh} = \Pi R^2 \int \rho g\, z(x) = \rho g\, (2/3)\, \Pi R^3$

Knowing that the equation of the circle is : $z(x) = \sqrt{R^2 - x^2}$ and that its derivative is : $z'(x) = x/z(x)$, we can rewrite (36) as :

$$\Sigma^{H}_{Res}(z) = -\rho g\,\lambda\, e^{-\varepsilon z} + \rho g\,\beta\, e^{-\alpha z} + \rho g\, x(z) \qquad (37)$$

Note that the stress gradient graph is of no use in calculating the parameters, because equilibrium (at $R=R_{ch}$) cannot be calculated at the level of gradients but at the level of forces.
Indeed, it is the balance of forces, calculated by integrating gradients, that allows us to calculate the solid-liquid parameter 'β', knowing the critical radius of the hemisphere '$R_{ch}$'.

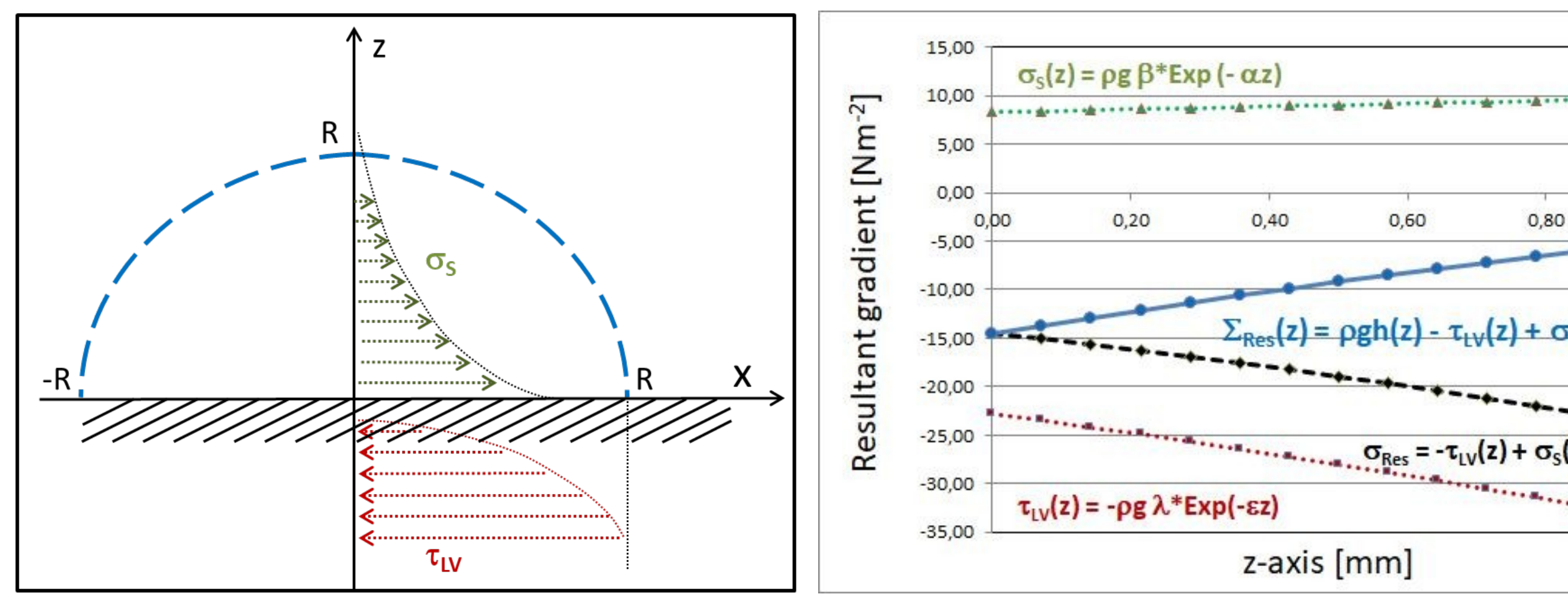

**Figure 15a – Scheme of gradients of horizontal compression and dilatation forces in a hemispherical drop**
**Figure 15b – Graph of the resulting gradient : $\Sigma(z) = \sigma_S(z) - \tau_{LV}(z) + \rho g\, x(z)$**
**Using following parameters : R= 1 mm ; λ = 4 mm, ε = 409 [$m^{-1}$], α = 167 [$m^{-1}$], β = 1 mm**

### 5.3.4 Equilibrium of the horizontal forces in the hemispherical drop

Surface tension horizontal forces are calculated by integrating the above gradients.
In the case of a hemispherical drop, the following inequality can be written as :

$$2\Pi R \int -\rho g\,\lambda\, e^{-\varepsilon z} + 2\Pi R \int \rho g\,\beta\, e^{-\alpha z} + \Pi R^2 \int \rho g\, x(z) \le 0 \qquad (38)$$

Where : '$\Pi R^2 \int \rho g\, x(z)$' is the hemispherical drop weight, i.e. : $\rho g\,(2/3)\,\Pi R^3$.
NB : even if we have a hemisphere, the total pressure is calculated by integrating over the perimeter '$2\Pi R$' as in the case of the sphere, because there is a substrate response that balances the vertical compressive forces.
Which gives, after integration :

$$\lambda/\varepsilon\,(1 - e^{-\varepsilon R}) - \beta/\alpha\,(1 - e^{-a R}) \le (1/3)\,R^2 \qquad (39)$$

This means that as long as the radius 'R' remains smaller than the critical radius '$R_{ch}$', the drop will theoretically remain hemispherical. When 'R' is greater than '$R_{ch}$', the hemispherical drop transforms into a greater non-hemispherical drop, e.g. semi-ellipsoidal, in order to achieve a new balance.
Of course, when the radius is equal to the critical radius ($R = R_{ch}$), inequality (39) becomes an equation.

### 5.3.5 Estimation of 'β' and of the critical radius '$R_{ch}$'

By analogy with what we did in the case of the spherical drop with equation (29), we should be able to calculate 'β', knowing the value of '$R_{ch}$'.
However, since there is no information about the value of the critical radius for a hemispherical drop, it can only be estimated by setting the value of 'β' and writing the balance between surface tension forces and drop weight as in equation (40) represented **Figure 16** :

$$\lambda/\varepsilon\,(1- e^{-\varepsilon R_{ch}}) - \beta/\alpha\,(1- e^{-\alpha R_{ch}}) = (1/3)\,R_{ch}^2 \qquad (40)$$

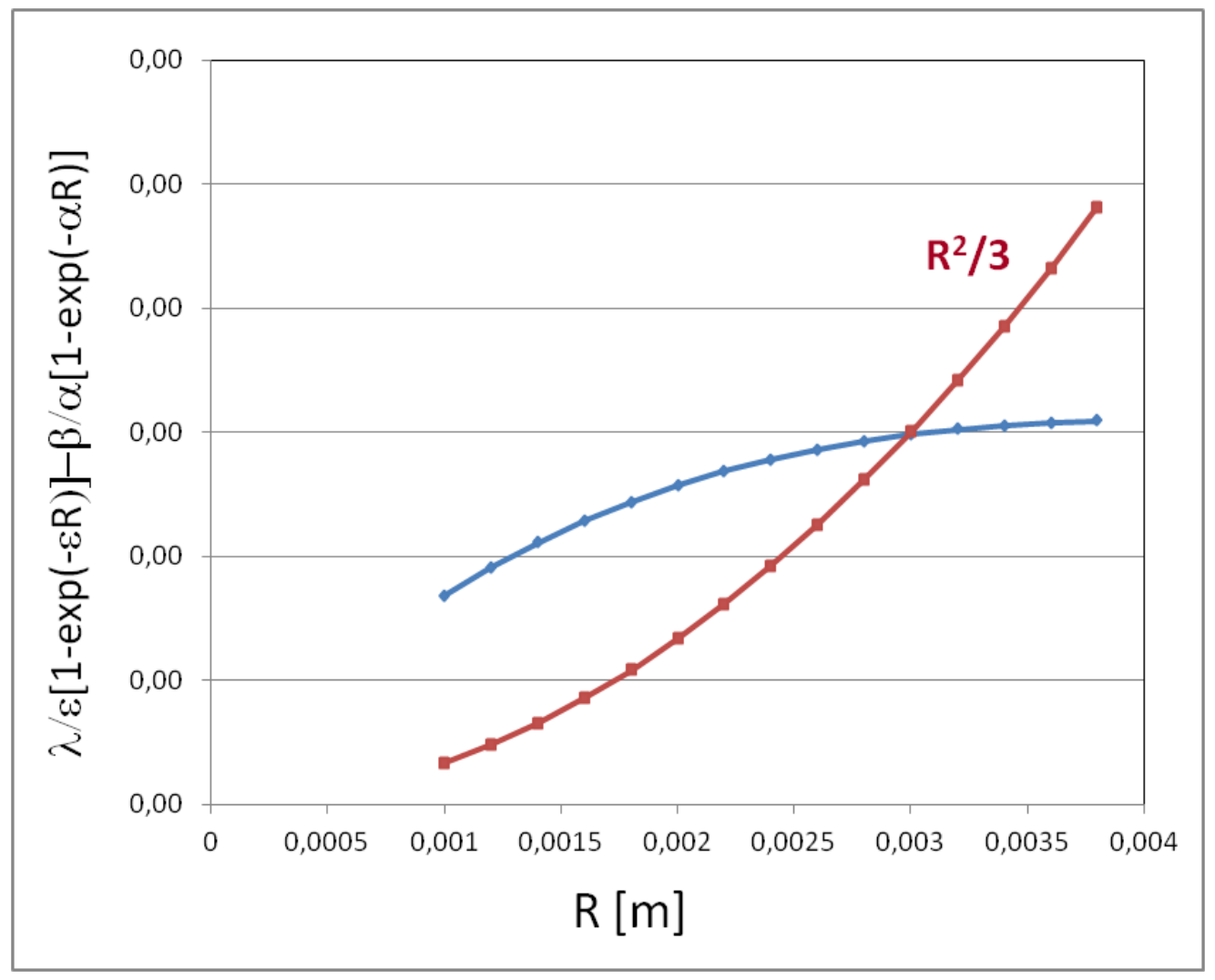

**Figure 16 – Graph of equation (40) with the following parameters estimated for water : λ= 4 mm, ε = 409 [$m^{-1}$], α = 167 [$m^{-1}$]**

For example, by estimating the value of 'β' to be : β = 1 mm , we get : $R_{ch}$ = 3,84 mm.
Remember that for a spherical drop, when setting the parameters : λ = 4 mm and ε = 409 [$m^{-1}$], we got : $R_c$ = 3,3 mm.
These parameters could be determined more precisely by means of experimental measurements.

### 5.3.6 Calculating the pressure variation in the hemispherical drop

The pressure in the hemispherical drop is calculated as :

$$\Delta P = \rho g\,\{2\Pi R \int - \lambda\, e^{-\varepsilon z} + 2\Pi R \int \beta\, e^{-\alpha z}\} / \Pi R^2 \qquad (41)$$

Then, after integration :

$$\Delta P = (2\rho g/R)\,\{\lambda/\varepsilon\,(1 - e^{-\varepsilon R}) - \beta/\alpha\,(1 - e^{-\alpha R})\} \qquad (42)$$

This equation can be compared to that of the pressure in the sphere (32, 33), that is close to that of Laplace.

### 5.4 Application of our approach to the case of a semi-ellipsoid drop

The semi-ellipsoid drop case is often used in literature to describe a wetting drop, i.e. a flattened drop that spreads more or less widely due to its size, such as the one shown in **Figure 17**.
We will study here the case of a semi-ellipsoid drop with a circular base, called ellipsoid of revolution.
The standard equation for a semi-ellipsoid of revolution of diameter '2L' and of height 'H' is as follows :
$x^2/L^2 + z^2/L^2 + y^2/H^2 = 1$.
The volume is given by : $V_{se} = 2/3\ \Pi\ HL^2$.

Boundary conditions :

- The height 'H' of an ellipsoid is, by definition, less than the base radius 'L' : $H < L$.
- Note that when '$H > L$', we move to a pseudo-spherical shape model (which will be studied below)
- The maximum drop height should not exceed a critical height which will be noted '$H_c$'.

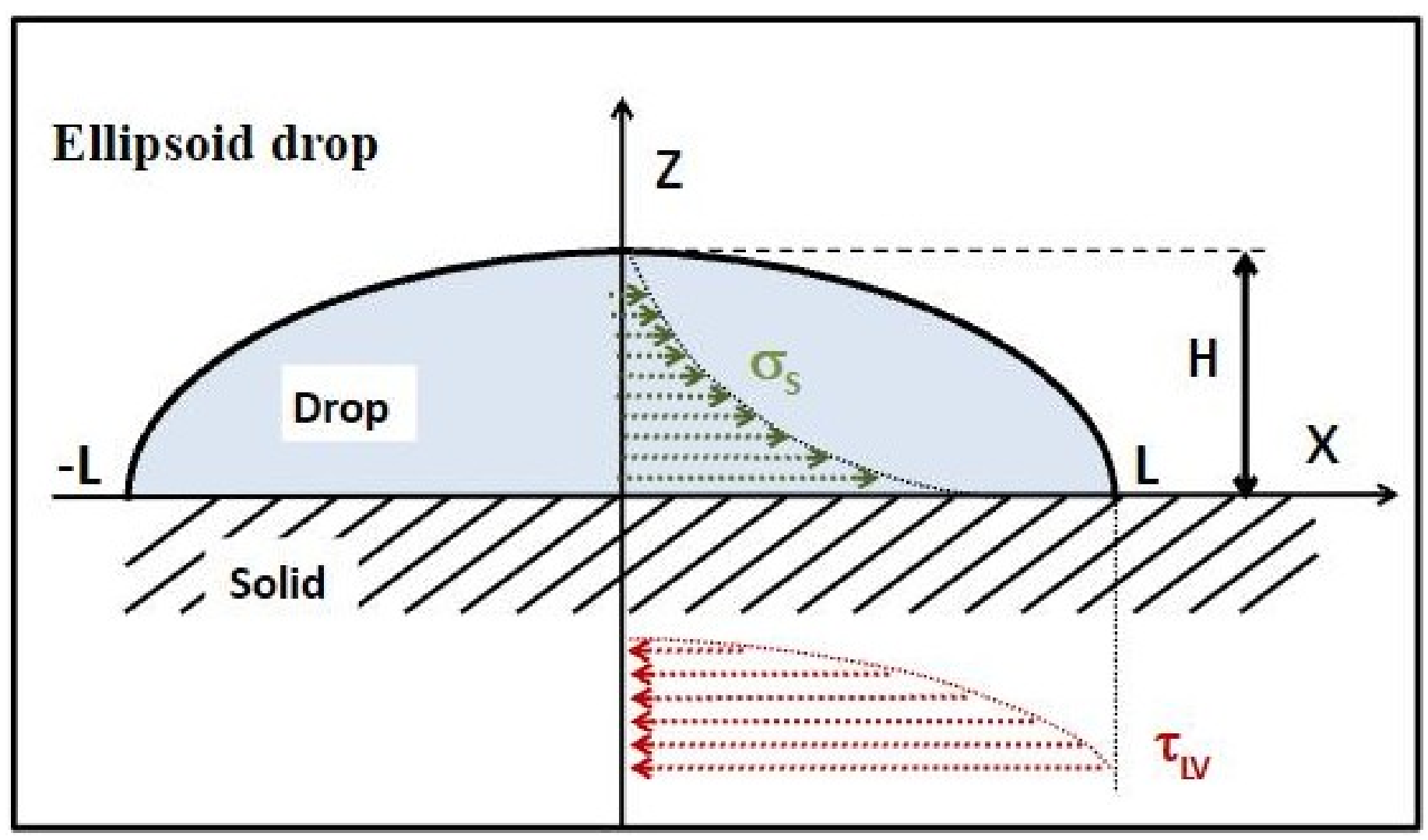

**Figure 17 – Schematic representation of the horizontal stress gradients in a semi-ellipsoid drop.**

#### 5.4.1 Critical radius and critical height concept for a semi-ellipsoid drop

While an ellipsoid of revolution has two dimensions, height and base radius, one might wonder if there is a critical radius or rather a critical height.
Note that the literature refers to semi-ellipsoidal drops of approximately 10 to 15 mm in diameter. Authors talk about 'dimensions beyond capillary length', such as those measured by F. Elie[23-24].
Therefore, it can be assumed that the critical radius of a semi-ellipsoidal drop should be greater than 5 mm.
This can be seen in **Figure 13**, where we see that there are large semi-ellipsoidal drops, which look a bit like puddles. These drops are not destroyed beyond a certain radius, unlike what is observed in the case of a spherical drop (the case of raindrops, for example).
In contrat, these drops have a height limit, suggesting that even if there is no critical radius, there may be a critical height that we will call '$H_c$'.
We will see below that this is what our equations lead to.

#### 5.4.2 Calculating stress gradients resultant in the semi-ellipsoid drop

The resultant of the horizontal stress gradients in the semi-ellipsoid drop is given by :

$$\Sigma^{H}_{Res}(x) = -\ \tau_{LV}(x) + \sigma_{S}(x) + \rho g\ z(x) \qquad (43)$$

Where, this time, 'z(x)' is the semi-ellipsoid profile of the drop.

The theoretical ellipsoid profile in a plan 'x0z' being given by : $x^2/L^2 + z^2/H^2 = 1$, the semi-ellipsoid drop profile can be written as : $z(x) = (H/L)\sqrt{L^2 - x^2}$

As previously, '$\Pi R^2 \int \rho g\, x(z)$' is the semi-ellipsoid drop weight, i.e. : $\rho g\, 2/3\, \Pi\, HL^2$.

Equation (43) can be written as :

$$\Sigma^{H}_{Res}(x) = -\,\rho g\, \lambda\, e^{-\,\varepsilon\,(L-x)} + \rho g\, \beta\, e^{-\,\alpha\,(L-x)} + \rho g\, (H/L)\sqrt{L^2 - x^2} \qquad (44)$$

As we did for the hemisphere, we will not plot the stress gradients since it is not useful for calculating the parameters, knowing that equilibrium is not calculated in terms of gradients, but in terms of forces.

### 5.4.3 Calculating the balance of forces in the semi-ellipsoid drop

As shown in **Figure 17**, horizontal forces act on the circumference '$2\Pi L$' and we can write :

$$2\Pi L\, \{ \int -\,\rho g\, \lambda\, e^{-\,\varepsilon\,(L-x)} + \int \rho g\, \beta\, e^{-\,\alpha\,(L-x)} \} + \rho g\, 2/3\, \Pi\, HL^2 \leq 0 \qquad (45)$$

NB : As in the case of the hemisphere, the integration is done on the semi-ellipsoid drop perimeter '$2\Pi L$', because there is a substrate response that balances the vertical compressive forces.

After integration, inequality (45) becomes :

$$H \leq 3/L\, \{(\lambda/\varepsilon\, (1 - e^{-\,\varepsilon\, L}) - \beta/\alpha\, (1 - e^{-\,a\, L})\} \qquad (46)$$

We can verify for the specific case of the hemisphere, i.e. when '$L = H$', that we find again the hemisphere equation (40) by substituting the base radius 'L' and the height 'H' by the hemisphere radius 'R'.

We have drawn the diagram of equation (46) in **Figure 18**.

- As assumed above, it appears that there is no critical radius, but rather a critical height '$H_c$'.
- The calculated line represents the critical drop height value '$H_c$' as a function of the semi-ellipsoid drop base radius 'L'. It corresponds to the balance of the surface tension forces with the weight, i.e. :

$$H_c = 3/L\, \{(\lambda/\varepsilon\, (1 - e^{-\,\varepsilon\, L}) - \beta/\alpha\, (1 - e^{-\,a\, L})\} \qquad (47)$$

Remarks :

- The curve of equation (47) stops at '$H=L$', because we find again the hemisphere equation with a critical radius of about 3,8 mm.
- We have drawn the area where : '$H>L$' with dotted lines, as it is related to the pseudo-sphere case that will be analyzed later.
- Inequality (46) corresponds to the hatched area under the curve of equation (47). For a given value of the base radius 'L', inequality (46) corresponds to any value of '$H \leq H_c$'.

In this diagram, the hemispherical drop is represented by the straight line '$H=L$' which ends at : $L = Rc$.

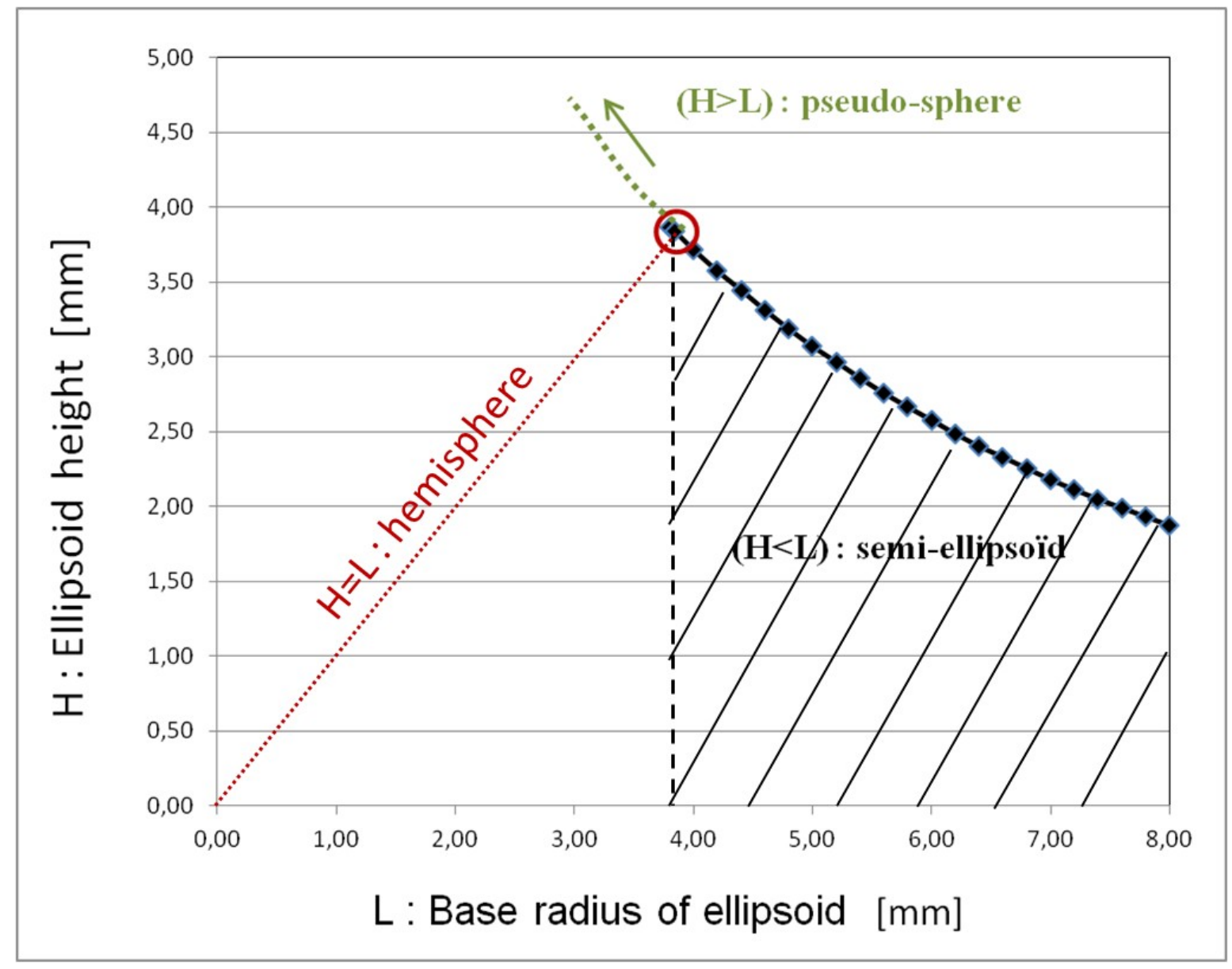

**Figure 18 – Graph of the semi-ellipsoid critical height '$H_c$' as a function of the base radius 'L' according to equation (46). For a given 'L' radius, 'H' values are smaller than '$H_c$'**

### 5.4.4 Calculating the pressure variation in the semi-ellipsoid drop

The semi-ellipsoid pressure drop is calculated on the surface '$\Pi L^2$' as :

$$\Delta P = \rho g \; \{ 2\Pi L \int_0^H - \rho g \, \lambda \, e^{-\varepsilon z} + 2\Pi L \int_0^H \rho g \, \beta \, e^{-\alpha z} \} / \Pi L^2 \tag{48}$$

Namely :

$$\Delta P = 2\rho g/L \; \{(\lambda/\varepsilon \, (1 - e^{-\varepsilon L}) - \beta/\alpha \, (1 - e^{-a L})\} \tag{49}$$

This equation can be compared to that of the pressure in the sphere (33) and in the hemisphere (42).
To clearly identify the theoretical values authorized by our model, we must now trace the other drop shape, namely the pseudo-spherical drop.

### 5.5 Applying of our approach to the case of a pseudo-spherical drop

The case of the pseudo-spherical drop described in **Figure 19** corresponds to the case described in the literature as that of a drop with low 'wetting' properties, i.e., a drop with a relatively low solid-liquid surface tension '$\gamma_{SL}$' (or a drop with a relatively high liquid-vapor surface tension '$\gamma_{LV}$'),
The volume of a pseudo-spherical drop '$V_{ps}$' is calculated as that of a sphere from which a spherical dome of height '2R-H' has been removed, namely : $V_{ps} = (4/3)\Pi R^3 - \Pi \, (2R-H)^2 \, (R+H)/3$.
As for the hemisphere and for the hemi-ellipsoid, the horizontal forces in the pseudo-sphere are calculated by integrating the gradients '$\tau_{LV}(z)$'et '$\sigma_S(z)$.

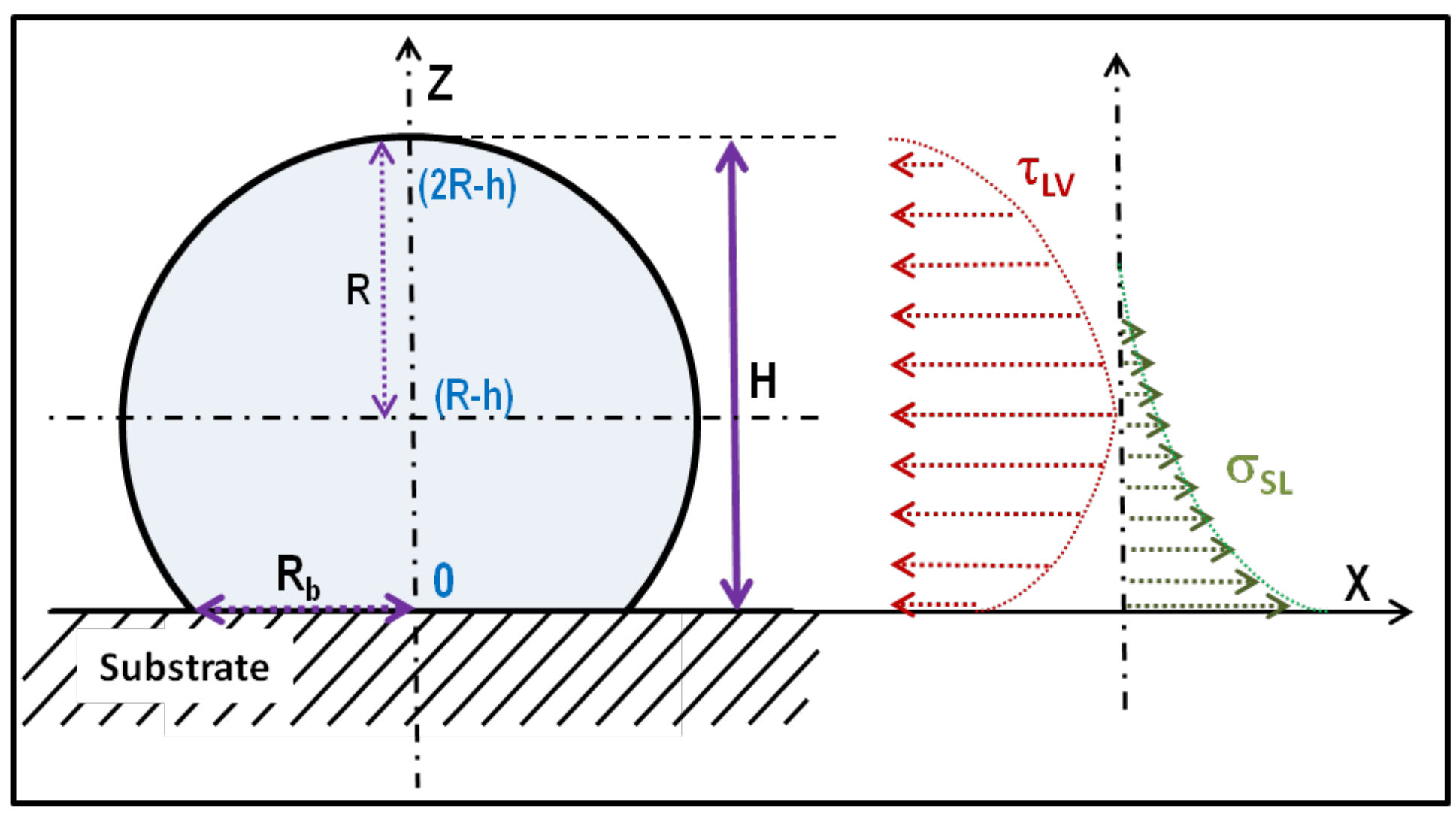

**Figure 19 – Schematic representation of the horizontal stress gradients in a pseudo-spherical drop.**

### 5.5.1 Calculating the resultant stress gradients in the pseudo-spherical drop

Preliminary assumption : We assume that the substrate response to vertical compressive forces is equal and opposite to the contribution of the spherical dome.
The horizontal stress gradients in the pseudo-spherical drop are calculated as follows :

- $\sigma_S(z)$ : is the resulting solid interface stress gradient
    - It integrates from z = 0 to z = H on the perimeter section '$2\Pi R_b$'
- $\tau_{LV}(z)$ : is the liquid-vapor interface stress gradient, negative because the drop is in compression
    - It integrates from z = 0 to z = R on the perimeter '$2\Pi R$' as a whole sphere from which we subtract the contribution of a spherical dome
- $\tau^{ca}{}_{LV}(z)$ : is the stress gradient in the spherical dome. It subtracts from the liquid-vapor interface stress gradient, and it is calculated on the perimeter section '$2\Pi R_b$'
- R(z) : is the delta response of the substrate to vertical compressive forces. It is negative and it also integrates on the perimeter section '$2\Pi R_b$'. It opposes the stress gradient of the spherical dome, and it will therefore cancel it out.
- $E_{pp}$ : is the hydrostatic pressure gradient (or potential gravity energy per unit volume)
    - It is exerted vertically, and its horizontal component is obtained as before
    - It helps to calculate the weight as : $P_{gps} = \rho g \, \{(4/3)\Pi R^3 - \Pi \, (2R-H)^2 (R+H)/3\}$

As for hemisphere and for hemi-ellipsoid cases, it is useless to draw the stress gradients graph because the equilibrium is not calculated at the level of gradients but at the level of the forces.

### 5.5.2 Calculating the balance of forces in the pseudo-spherical drop

The resultant of the forces can be calculated by integrating the following three gradients: the liquid-vapor stress gradient at the equatorial perimeter of the drop ($2\Pi R$), the solid-liquid stress gradient and the spherical dome stress gradient at the base ($2\Pi R_b$). The inequation is then written as :

$$2\Pi R_b \int_0^H \rho g \beta \, e^{-\alpha z} - 2\Pi R \int_0^R \rho g \, \lambda \, e^{-\varepsilon z} - \{ 2\Pi R_b \int_0^{H-2R} \tau^{ca}{}_{LV}(z) + 2\Pi R_b \int_0^{H-2R} R(z) \}$$
$$+ \rho g \, [(4/3)\Pi R^3 - \Pi(2R-H)^2 (R+H)/3] \leq 0 \qquad (50)$$

That is :

$$2\Pi R_b \int_0^H \rho g \beta\, e^{-\alpha z} - 2\Pi R \int_0^R \rho g\, \lambda\, e^{-\varepsilon z} + \rho g\, \{(4/3)\Pi R^3 - \Pi(2R-H)^2 (R+H)/3\} \leq 0$$

We get :

$$R\, \lambda/\varepsilon\, (1 - e^{-\varepsilon R}) - R_b\, \beta/\alpha\, (1 - e^{-a H}) \leq (2/3)R^3 - (2R-H)^2 (R+H)/6 \quad (51)$$

Note that the contribution of the stress gradient of the spherical cap has been removed by assuming that the substrate response to the compressive forces is equal and opposite to this contribution (which leads to the disappearance of the terms in the brackets '{}' in (51)).

This hypothesis can be easily verified using the following two special cases : the case of a spherical drop (when : $R_b = 0$) and the case of a hemispherical drop (when : $R_b = H$) :

- Case of a spherical drop ($R_b = 0$)

When : $R_b = 0$, we get : H = 2R and equation (51) returns to equations (29, 30) obtained for the sphere, where the equilibrium of the resulting forces with the weight is reached when the radius is equal to the critical radius. '$R_c$'.

- Case of a hemispherical drop ($R_b = H$)

When : $R_b = H$, we get : H = R and equation (51) returns to equation (40) obtained for the hemisphere where the balance of the resulting forces leads to inequality. Let us recall that in the case of the hemispherical drop, the equilibrium is reached when the radius is equal to the critical radius '$R_{ch}$'. When the radius is greater than the critical radius, the hemispherical drop changes to a larger drop, for example a semi-ellipsoid.

As shown in **Figure 20**, we can now use inequation (51) to plot the height diagram 'H' as a function of the base radius '$R_b$' for the pseudo-spherical drop. Inequation (51) has been merged with inequation (46) which gives the semi-ellipsoidal drop height as a function of the radius 'L'. We can use it to represent the cases of the sphere (no contact) and the hemisphere and draw the diagram of the diverse types of drops deposited on a substrate.
The interpretation of this graph has been set out in paragraph 5.6.

### 5.5.3 Calculating the pressure difference in the pseudo-spherical drop

The pseudo-spherical drop pressure can be calculated on the surface '$\Pi R^2$' using (50) as :

$$\Delta P = \{\, 2\Pi R_b \int_0^H \rho g \beta\, e^{-\alpha z} - 2\Pi R \int_0^R \rho g\, \lambda\, e^{-\varepsilon z} \,\} / \Pi R^2 \quad (52)$$

with : $R_b = \sqrt{H(2R-H)}$

Namely :

$$\Delta P = (2\rho g/R)\, \{\lambda/\varepsilon\, (1 - e^{-\varepsilon R}) - (R_b/R)\, \beta/\alpha\, (1 - e^{-a H})\} \quad (53)$$

We see that this equation slightly differs from the semi-ellipsoid by a factor of '$(R_b/R)$'.

## 5.6 Diagram of different liquid drops deposited on a solid surface

We can now collect the inequations of the different drop patterns defined above and draw the diagram of the different drop patterns deposited on a solid surface as in **Figure 20**. We have merged the inequalities (29, 39, 46 et 51) to plot the height of the drops 'H' as a function of the radius 'L' of semi-ellipsoidal drops and the base radius 'Rb' of pseudo-spherical drops.

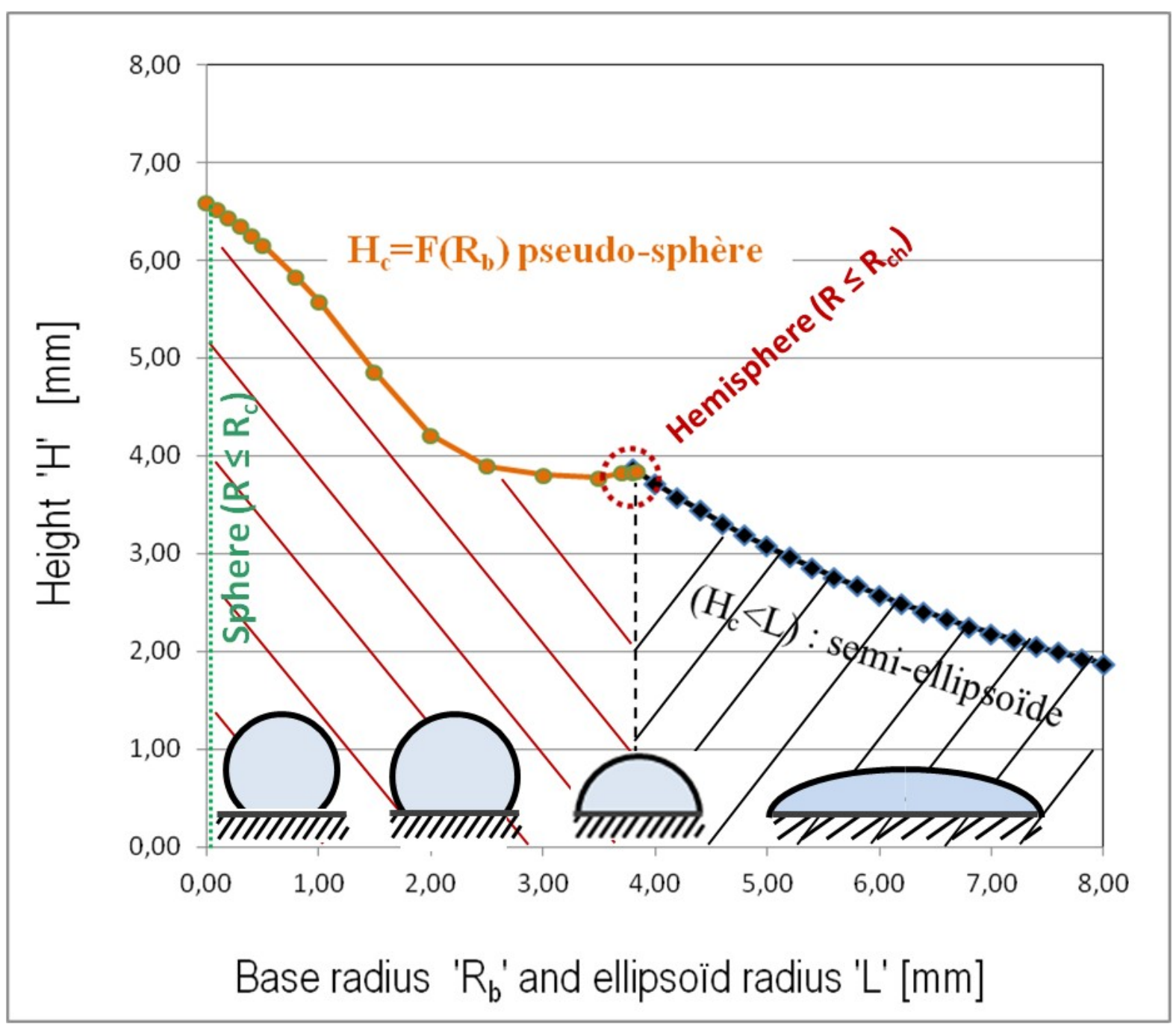

**Figure 20 – Height diagram of different drop shapes as a function of the semi-ellipsoidal drop base radius 'L' according to (46) and as a function of the pseudo-spherical drop base radius '$R_b$' according to (51)**

This diagram can be interpreted as follows :

- For a semi-ellipsoidal drop, inequation (46) corresponds to any value of height '$H \leq H_c$' for a given radius 'L'. It corresponds to the shaded area under the curve of equation (47).
- The hemispherical drop is represented by a straight line '$H = L$' that stops at '$L = R_{ch}$'.
- For a perfect spherical drop, we can use inequation (30) by replacing 'H' with '2R' (provided that : $R \leq R_c$).
- For a pseudo-spherical drop, inequation (51) corresponds, for a given base radius '$R_c$', to any value of height '$H \leq H_c$' in the shaded area under the curve.

Regarding the shape of the drops, knowing that by definition the size of a sphere is limited by '$R_c$' and the size of a hemisphere is limited by '$R_{ch}$', it means that for drops of radius 'L' greater than '$R_{ch}$', we should have ellipsoids.
On the other hand, nothing prevents a small drop from having an ellipsoidal shape.

So, according to our equations, we should theoretically observe small drops of any shape, but we should not theoretically observe large hemispherical drops nor large pseudo-spherical drops if the radius is greater than '$R_{ch}$'. According to photographs seen in the literature, it seems that small and medium sized drops often have a pseudo-spherical or other shape, while larger drops often have a semi-ellipsoidal shape, which confirms our analysis.

## VI Relationship between surface stress gradients and surface tension : definition of an equivalent term ‘Γ’

Inevitably, the question arises whether we can relate our equations, and in particular the previously defined term ‘Γ’ to the value of the classical surface tension ‘γ’.
To explore this link, we need to establish a few preliminary definitions.

### 6.1 Preliminary definitions

In section 3.2, we previously defined ‘Γ’ as an equivalent term to the classically defined surface tension ‘γ’, both terms having the same dimension in [$Nm^{-1}$].
Indeed, the term ‘Γ’ is obtained by integrating the stress gradient in the normal direction to the surface and, since the gradient is expressed as a force per unit area, its integral is then expressed as a force per unit length.

The terms ‘$\Gamma_S$’ and ‘$\Gamma_L$’ can be defined as follows :

- The equivalent liquid-vapor term ‘$\Gamma_L$’ corresponds to liquid-vapor surface tension ‘$\gamma_{LV}$’ such as :

$$\Gamma_L = \int \tau(z)\, dz \quad [Nm^{-1}] \tag{54}$$

- The equivalent solid-liquid term ‘$\Gamma_S$’ corresponds to the resultant of the solid-liquid surface tension ‘$\gamma_{SL}$’ and the solid-vapor surface tension ‘$\gamma_{SV}$’ as follows:

$$\Gamma_S = \int \sigma(x)\, dx \quad [Nm^{-1}] \tag{55}$$

### 6.2 Comparison of equivalent terms ‘Γ’ to surface tensions ‘γ’

Let us compare the terms ‘Γ’ and ‘γ’ in the following four cases.

(i) Let us start with the generic case of an extended water surface, which allows for the comparison of the equivalent term ‘$\Gamma_L$’ with the classical surface tension ‘$\gamma_{LV}$’.

In the case of an extended water surface, such as the one shown in **Figure 10a**, surface tension forces cannot bend the surface because the volume is large. The equivalent term ‘$\Gamma_L$’ is obtained by integration of the surface stress gradient ‘$\tau_{LV}(z)$ over an interval ranging from zero to infinity (i.e., at a distance greater than the attenuation distance ‘W’) such as :

$$\Gamma_L = \int \tau_{LV}(z)\, dz = \int \rho\ g\ \lambda\ e^{-\varepsilon z} = \rho\, g\, \lambda / \varepsilon \quad [Nm^{-1}] \tag{56}$$

Using the previously estimated values of the parameters ‘ε’ and ‘λ’ for water, a generic value of ‘$\Gamma_L$’ of $72.10^{-3}$ [$Nm^{-1}$] is obtained, identical to the classical value given in the literature for the surface tension ‘$\gamma_{LV}$’.
Thus, in the case of an extended water surface, the equivalent term ‘$\Gamma_L$’ has de same value as the surface tension ‘$\gamma_{LV}$’.

(ii) In the case of the meniscus as shown in **Figure 4a**, the surperficial tension forces and gravity are in equilibrium. Using equation (12), the resulting term '$\Gamma_R$' can be related to the resulting stress gradient '$\sigma_R(x)$' such as :

$$\Gamma_R = \int \sigma_R(x)\, dx = \int_0^W \rho\ g\ A\ e^{-Bx}\, dx \qquad (57)$$

Considering that the meniscus is free to extend to infinite, the integration yields :

$$\Gamma_R = \rho\, g\, B/A \qquad (58)$$

In paragraph 3.2, it was linked to the product '$\gamma_{LV} . \cos(\theta)$' commonly measured by experimenters.
Remember that the term '$\Gamma_R$' is considered as a combination of '$\Gamma_S$' and '$\Gamma_L$'. It is this resulting term that is obtained, for example, in the measurements made with the Wilhelmy Tensiometer and the capillary tube case (Jurin's law).
Using the values of parameters 'A' and 'B' previously estimated for water (A = 316 [$m^{-1}$] and B = $2.10^{-3}$ [m]), a value '$\Gamma_L$' of $62.10^{-3}$ [$Nm^{-1}$] is obtained. This value corresponds exactly to the value given in the literature (at 25°C) for surface tension with an angle of 30 degrees ($72.10^{-3}$ . cos(30°) [N/m]).
Thus, in the meniscus case, the equivalent term '$\Gamma_R$' has the same value as the product '$\gamma_{LV}$' . cos($\theta$)'.

(iii) In the case of a spherical isolated drop as shown in **Figure 10a**, surface tension forces will bend the surface because the volume is small compared to the attenuation distance '$W_g$'.
Apart from gravity, the only forces that come into play are the liquid-vapor interaction forces and the equivalent term '$\Gamma_L$' can be calculated by integrating the liquid-vapor stress gradient '$\tau_{SL}(r)$' using equation (26) as :

$$\Gamma_L = \int \tau_{SL}(r)\, dr = \int_0^R \rho\ g\ \lambda\ e^{-\mathcal{E}r}\, dr \qquad (59)$$

Integration gives :

$$\Gamma_L = \rho\, g\, \lambda/\mathcal{E}\, (1 - e^{-\mathcal{E}R}) \qquad (60)$$

In the gravitational field, the drop radius 'R' is, by definition, smaller than the critical radius '$R_c$' and much smaller than '$W_g$'. One can relate '$\Gamma_L$' to the classical surface tension '$\gamma_{LV}$' by calculating the drop pressure as in equation (31), and one can compare this equation to Laplace's equation ($\Delta P = \frac{2\ \gamma}{R}$) such as :

$$\Delta P = \frac{2}{R}\, \rho\, g\, \lambda/\mathcal{E}\, (1 - e^{-\mathcal{E}R}) = \frac{2}{R}\, \Gamma_L \qquad (61)$$

Where '$\Gamma_L$' is equivalent to '$\gamma_{LV}$' and can be written as : $\Gamma_L \cong \rho\, g\, \lambda/\mathcal{E}\, (1 - e^{-\mathcal{E}R})$ [$Nm^{-1}$].
Using the values of parameters '$\varepsilon$' and '$\lambda$' estimated earlier for water and with a maximum radius of 4 mm, a value '$\Gamma_L$' of $77.10^{-3}$ [$Nm^{-1}$] is obtained, close to the value of '$\gamma_{LV}$'.

Note that, that in zero gravity we should have, as in equation (56) : $\Gamma_L \cong \rho\, g\, \lambda/\mathcal{E}$. In this case, we would obtain '$\Gamma_L$' : $72.10^{-3}$ [$Nm^{-1}$], close to the value of '$\gamma_{LV}$' given in the literature.

Thus, in the case of a spherical isolated drop, the equivalent term '$\Gamma_L$' is close to the value of the classical surface tension '$\gamma_{LV}$'.

(iv) In the case of drops deposited on a plane surface as shown in **Figures 14, 17** and **19**, we have seen that the conjunction of solid and liquid-vapor surface tension forces leads to the formation of drops whose shape varies from a semi-ellipsoid (wetting drop) to a pseudo-sphere (non wetting drop).

In the hemispherical drop case of **Figure 14**, '$\Gamma_S$' and '$\Gamma_L$' terms can be calculated and introduced into the pressure equation (42) as follows :

$$\Delta P = (2/R) \{ \Gamma_L - \Gamma_S \} \quad (62)$$

Where :

$$\Gamma_L = \rho g\, (\lambda/\varepsilon)\, (1 - e^{-\varepsilon R})$$
$$\Gamma_S = \rho g\, (\beta/\alpha)\, (1 - e^{-\alpha R})$$

We can also use these terms to rewrite the balance of forces in the hemispherical drop of equation (39) :

$$(\Gamma_L - \Gamma_S) \leq (1/3)\, \rho g\, R^2 \quad (63)$$

At this point, the two theories diverge : clearly, our equation (63) cannot be compared with that of Young-Dupré (2), because the two approaches are different.
Thus, after observing a convergence in the first three cases, we observe a fundamental divergence in the latter case, which we propose to discuss in the following paragraph.

### 6.3 Interpretation of divergences between equivalent terms '$\Gamma$' and surface stresses '$\gamma$'

In the case of a hemispherical drop as shown in **Figures 14** and **15a**, our mechanical equivalence model considers only three forces :

- The liquid-vapor interaction forces normal to the drop surface, which are compressive forces
- The resultant of solid-liquid and solid-vapor interaction forces that cause the drop to spread out
- The force of gravity, which opposes the resultant of the superficial tension forces and also tends to crush the drop.

Using **Figure 21**, we can compare our view (**Figure 15a**) with that of Young-Dupré (**Figure 1**) :

(1) The liquid-vapor force gradient '$\tau_{LV}$' is oriented toward the center of the drop, just like the projection of the vector '$\gamma_{LV}$'.
(2) The gradient resulting from the solid-liquid and solid-vapor forces '$\sigma_S$' runs from the center to the edge of the drop, whereas, in Young-Dupré's law, the vectors '$\gamma_{SL}$' and '$\gamma_{SV}$' are opposite : The vector '$\gamma_{SL}$' goes from the edge to the center of the drop, while the vector '$\gamma_{SV}$' goes toward the outside of the drop. This is an apparent difference because, it should be remembered that we have defined '$\sigma_S$' as the resultant of the solid-liquid and solid-vapor forces, and its theoretical components '$\sigma_{SL}$' and '$\sigma_{SV}$' are probably one negative and the other positive, like '$\gamma_{SL}$' and '$\gamma_{SV}$'.
(3) First difference : In the case of an isolated drop such as the one in **Figure 9b**, we consider the term '$\Gamma_L$' to be radial, while the surface tension '$\gamma_{LV}$' is considered to be tangential.
(4) Second difference : The Young-Dupré equation does not consider the effect of gravity, whereas we believe that gravity has a significant effect on the cohesion of the drop (notion of critical radius) as well as on the rise of the meniscus.

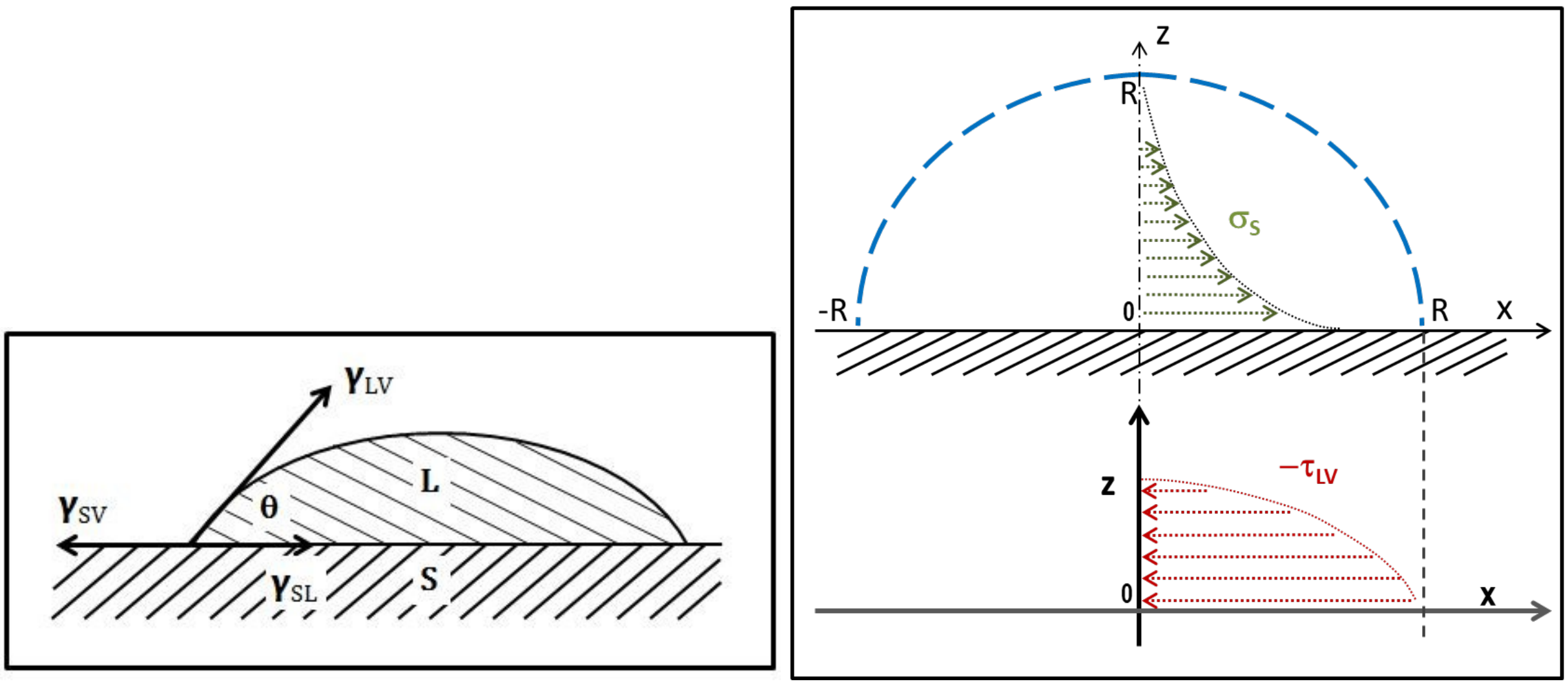

**Figure 21 – Comparison of force vector schemes in a hemispherical drop according to the Young-Dupré equation and the stress gradient according to this paper**

In summary, the approaches are different but describe the same phenomena, with the differences being due to a vision of tangential forces per unit length in the case of Young-Dupré's law, whereas our model uses force gradients per unit area that are expressed in volume.
Therefore, although the equivalent terms 'Γ' and surface stresses 'γ' have the same dimension and similar values when calculating the equilibrium of forces, the approaches are fundamentally different.
The conclusion is clear : defining equivalent terms to bring our equations closer to classical equations does not seem particularly useful.

## VII DISCUSSION

In this article, we propose a thought experiment that consists in mathematically replacing the surface tension by an equivalent stress gradient exerted in volume, below the surface.
More specifically, rather than using the classic surface tension in [N/m], we hypothesize that the attraction or repulsion of molecules at the interface creates a gradient of deformation or compression stress perpendicular to the surface in [N/m²]. These force gradients act within the volume of the liquid and over a long distance, explaining, for example, the size of the meniscus. Thus, rather than considering that surface energy is concentrated at the interface on a mathematical surface with no thickness, we consider here that it is expressed in volume.
Although no experimental measurements have yet been made to validate such a hypothesis, we can imagine a few theoretical considerations that might explain such a gradient :
(i) The gradient could be linked to a local variation in viscosity during the formation of the meniscus, given that viscosity depends on the speed of the fluid and that viscosity slows down the evolution of the film.
(ii) The gradient could be a gradient of molecular reorganization that varies depending on the distance from the interface according to a stationary process of creation/destruction of short-lived structures could lead to a gradient of molecular reorganization that varies with distance from the interface. At the end of the previous formation of the meniscus (or the capillary rise), a stationary regime of creation/destruction could be established, a regime that would be independent of time[26] but dependent on the distance from the wall. In the case of water, such a reorganization gradient could be achieved by the creation/destruction of very short-lived structures using hydrogen bonds.
Regarding the range of action of such a molecular reorganization process, it should be noted that if L.J. Michot et al.[10] claim that the structural perturbations at the interfaces do not extend beyond distances greater than 10-15 Å from the interface, J.M. Zheng et al[27] do not fully agree with this viewpoint and indicate that they observed structural perturbations extending over several hundred microns, Indeed, these authors report that colloidal and molecular solutes suspended in an aqueous solution can be largely excluded from the vicinity of various hydrophilic surfaces over a distance of several hundred microns. Moreover, regarding the nature of the precursor film, Popescu et al.[24] reported that the precursor film is a "mesoscopic film whose propagation includes the contribution of long-range forces".

## VIII CONCLUSION

In this paper, we proposed a thought experiment in which the surface tension has been mathematically replaced by an equivalent stress gradient exerted below the surface.
Using such a gradient, we can rewrite the equations of surface tension in terms of force per unit area or energy per unit volume, whereas they are classically written in terms of force per unit length or energy per unit area.
We have temporarily developed a mechanical equivalence model that defines two stress gradients, a surface stress gradient at the liquid-vapor interface '$\tau_{LV}(x)$' and a resulting stress gradient at the solid interfaces '$\sigma_S(x)$', because this model does not allow for discrimination between the components '$\sigma_{SL}(x)$' and '$\sigma_{SV}(x)$'.
This model nevertheless allows us to reinterpret the previous equations that describe known phenomena (meniscus, capillary tube, Wilhelmy plate, and drops), to provide an improved expression of Laplace's law, to integrate the equation of gravity to explain the size limit of water drops on Earth, to explain their behavior in weightlessness and model the formation of hemispherical, semi-ellipsoidal, and pseudo-spherical drops.
We believe that developing a phenomenological model will allow us to consider all components, a topic that will be addressed in a future article.